\documentclass[11pt]{article}
\usepackage{amssymb}
\usepackage{amstext}
\newcommand*\de{\mathrm{d}}
\newcommand*\De{\mathrm{D}} 
\renewcommand*\epsilon{\varepsilon}
\renewcommand*\phi{\varphi}
\renewcommand*\theta{\vartheta}
\begin{document}
  
\title{\bf Generally covariant quantization and the Dirac field} 

\author{M. Leclerc \\ \small Section of Astrophysics and Astronomy, 
Department of Physics, \\ \small University of Athens, Greece}  
\date{June 7, 2006}
\maketitle 
\begin{abstract}
Canonical Hamiltonian field theory in curved spacetime 
is formulated in a manifestly covariant way. 
Second quantization is achieved invoking a  correspondence 
principle between the  Poisson bracket of classical fields 
and the commutator of the corresponding  quantum operators.
 The Dirac theory is investigated and 
it is shown that,  in  contrast to 
the case of bosonic fields, in curved spacetime, the field momentum 
 does not coincide with the generators of spacetime translations. 
The reason is  traced back to the presence of second class constraints 
occurring in Dirac theory. Further, it is shown that the modification of 
the Dirac Lagrangian by a surface term leads to a momentum transfer between 
the Dirac field and the gravitational background field, resulting in a 
 theory that is free of  constraints, but not manifestly hermitian. 
\end{abstract}

\section{Introduction}

Quantization in curved spacetime has a long history and there exists an 
equally long list of problems related to the subject. It is not our intention  
to give a review of those issues (see, e.g., \cite{dewitt} for a discussion 
of many of the related  problems as well as for a 
list of relevant references). 
Here, instead, we wish, in a certain sense,  to start from zero and 
 investigate some of the consequences of a straightforward 
canonical quantization     
performed in curved spacetime that arise independently of eventual  
additional 
problems like those described in \cite{dewitt} and many other articles 
(e.g., the observer independent concept of a particle). We base our 
investigation on the principle of relativity, which states that, locally, 
we cannot distinguish between gravitational and inertial fields. Therefore, 
if we can put a given special relativistic theory into a manifestly generally 
covariant form and perform, always in a manifestly covariant way, the second 
quantization, then the incorporation of gravitational background fields 
will be trivial, since all our relations will remain identical in form. 
(This holds as long as we stick to the minimal coupling principle, avoiding 
thus explicit curvature couplings.) According to this procedure, the 
special relativistic theory dictates the form of the corresponding 
theory in gravitational background fields. If the resulting theory  
is not free of problems, then there are essentially three possibilities: 
First, the problems might  be solved in some way. 
This is the point of view we adopt
here. In many cases, this involves difficulties concerning  the  
interpretation of certain results, which might be  straightforward in flat
spacetime, but leads to ambiguities in curved spacetime. Second, the 
canonical 
 quantization procedure might not be completely correct, and therefore 
leads to problems which, by chance, do not manifest themselves in the 
special relativistic limit. We do not investigate this possibility here. 
Finally, there is the possibility that it 
is simply not possible to perform second quantization in a manifestly
covariant  way, and thus, ultimately, in an  observer independent way. 
Otherwise stated, in such a 
case, the quantization process does not commute with the change of the 
coordinate system, i.e., we will necessarily have to fix first the 
coordinate system, and then perform the quantization. Obviously, this is 
not a very attractive option, since it would essentially 
mean that at the quantum level, 
the principle of relativity ceases to be valid. 

Assuming that the canonical quantization procedure is valid and can 
be performed in a generally covariant way, we proceed straightforwardly 
to the analysis of both bosonic and fermionic theories. We start by the 
formulation of classical field theory, where we use a direct 
generalization of the Poisson bracket used by Ozaki \cite{ozaki} 
in the framework of special relativistic theories, and perform the 
quantization of the theory by invoking the traditional correspondence 
principle between Poisson brackets and commutators (or anticommutators). 
The resulting quantum theory is essentially a curved spacetime 
generalization of Schwinger's 
manifestly Lorentz covariant formulation of special relativistic field 
theory \cite{schwinger1}.  

Of particular interest is the result that, in Dirac theory, the 
generators of spacetime translations are not given in terms 
of the field momentum operator. Thus, in particular, the time component of 
the field momentum, which is conventionally referred to as field Hamiltonian,  
 does not generate the time evolution of the quantum fields, as is the case 
in flat spacetime. Instead, it turns out that the field momentum 
generates a kind of generalized translations which are directly related 
to the hermitian momentum operators $\tilde p_k$ derived in our previous 
article \cite{leclerc1}. The reason is found in the occurrence of second class 
constraints in Dirac theory. 

Further, an alternative way to quantize the Dirac field is presented, 
where the Lagrangian is modified by a surface term. The surface term 
is shown to  lead to a change in  the field momentum, that can be interpreted 
as a momentum transfer between the Dirac field and the gravitational 
background field. The resulting theory is free of constraints, and 
the generators of spacetime translations are now given directly in terms 
of the field momentum operator. However, the theory is not manifestly 
hermitian, and the symmetry between the field variables $\psi$ and 
$\bar \psi$ is broken, the latter playing merely the role of a Lagrange 
multiplier. Most interesting, in the special relativistic limit, 
both the hermitian and the non-hermitian formulations 
turn out to be equivalent, which is the reason why the issue of how we 
deal consistently with the constraints in Dirac theory is usually  
passed over in the related literature.

The article is organized as follows. In the next section, we 
explain our notations and give a few definitions that will be used 
throughout this article. In sections \ref{poisson} and \ref{hamilton}
classical Hamiltonian field theory for bosonic fields 
is formulated in a manifestly covariant way, and in section \ref{second}, 
we proceed to second quantization. Finally, in section \ref{dirac}, 
we discuss the Dirac theory. 

\section{Preliminaries} 

Quite generally, our notations are identical to those used in
\cite{leclerc1}. In particular, we use latin letters  from the middle 
of the alphabet $i,k,l,m\dots$ to denote spacetime indices (e.g., 
the spacetime metric $g_{ik}$) and latin letters from the beginning of 
the alphabet $a,b,c\dots $ to denote Lorentz vector indices (e.g., 
the flat tangent space metric $\eta_{ab}$, the Lorentz (or spin) connection
$\Gamma^{ab}_{\ \ i}$, the tetrad field $e^a_i $, with $g_{ik} = e^a_i 
e^b_k \eta_{ab}$). Both spacetime and tangent space are four dimensional. 
At a later stage, we will also use spinor indices 
$L,M,N\dots$ (running from 1 to 4) and write the Dirac spinor as $\psi^M $
(as well as $\bar \psi_M$ for the conjugate spinor, transforming with the 
inverse under a Lorentz gauge transformation).

Let $x \equiv x^i = (x^0, x^1, x^2, x^3)$  be 
spacetime  coordinates such that a hypersurface element can be written 
as  
\begin{equation}\label{0} 
\de \sigma_i(x)  = \left( \begin{array}{c} 
\de x^1(\sigma) \de x^2(\sigma) \de x^3(\sigma) \\ 
 \de x^0(\sigma) \de x^2(\sigma) \de x^3(\sigma) \\  
\de  x^0(\sigma) \de x^1(\sigma) \de x^3(\sigma) \\ 
 \de x^0(\sigma) \de x^1(\sigma) \de x^2(\sigma)
\end{array}
\right), 
\end{equation}   
where  $\de x^i(\sigma) $ means that $\de x^i $ is restricted 
to some hypersurface $\sigma$ defined by $\Phi(x) = 0$. (E.g., for 
the hypersurface $x^0 = const$, we have $\de x^0 = 0$ and $\de \sigma_i(x) 
= \delta^0_i \de^3 x$.)

In the same coordinate system, we define 
\begin{equation} \label{1} 
\delta^i (x-y) = 
\left( \begin{array}{c}
\delta_{x^0 y^0}\  \delta(x^1-y^1)\delta(x^2-y^2)\delta(x^3-y^3)\\
\delta_{x^1 y^1}\  \delta(x^0-y^0)\delta(x^2-y^2)\delta(x^3-y^3)\\
\delta_{x^2 y^2}\  \delta(x^0-y^0)\delta(x^1-y^1)\delta(x^3-y^3)\\
\delta_{x^3 y^3}\  \delta(x^0-y^0)\delta(x^1-y^1)\delta(x^2-y^2)
\end{array}
\right).
\end{equation}

The transformation behavior for  $\delta^i (x-y)$ 
under a coordinate change is  found 
from the known transformation behavior of $\de \sigma_i(x)$ 
($\sqrt{-g}\ \de \sigma_i $ is a vector)  
by requiring $\delta^i (x-y) \de \sigma_i $ to transform as  scalar under 
general coordinate transformations.  Thus, $\delta^i(x-y)$ transforms as 
vector density.

Next, consider 
a spacelike hypersurface $\sigma$ defined by $\Phi(x) = 0$, with the  
 (timelike) normal 
 vector $n_i = \Phi_{,i}$. For convenience, $\Phi(x)$ can be chosen such 
that $n^2 \equiv n_{i}n_{k} g^{ik} = 1$. Then, we have 
\begin{equation} \label{2}
\int_{\sigma} f(x) \delta^i(x-y)  \de \sigma_i(x) = f(y) 
\end{equation}
where the integration is carried out over the  hypersurface $\sigma$ 
containing 
the point $y$. For the specific hypersurface $t = t_0=const$, we find, e.g., 
\begin{equation}  \label{3} 
\int_{\sigma} 
f(x) \delta^i(x-y)  \de \sigma_i(x) \!=\! \int_{\sigma} 
 f(t_x, \vec x) 
\delta^{(3)}(\vec x - \vec y)
\ \delta_{t_x t_y}\  \de^3 x 
\!=\! f(t_x, \vec y) \ \delta_{t_x t_y}, 
\end{equation}
where $t_x = t_0$ is to be taken on the hypersurface in question. Thus, if 
$t_y = t_0$ (i.e., if $y$ is on the hypersurface $t=t_0$), 
the result is simply 
$f(y)$, while else, we find zero. Thus,  $\delta^i (x-y) $ can be seen as 
 covariant generalization of the three dimensional delta function.

Note that we have adopted the convention of \cite{leclerc1} to use 
the index $t$ for the time component of a spacetime vector and 
greek indices from the middle of the alphabet $\mu,\nu \dots$ for 
the spacelike components, i.e., e.g., $A^m = (A^t, A^{\mu})= (A^t, 
\vec A)$, $\mu = 1,2,3$. 
The same letter $t$ is used for the time coordinate itself, , $x^m = 
(t, x^{\mu}) = (t, \vec x)$. If the time component of different events  
 $x^i, y^i$ is needed, we use the obvious notation $t_x, t_y$ etc.  
It is 
important to have in mind that, whenever such a $3+1$-split is 
used, it is understood that $t$ is really a timelike coordinate, i.e., in 
particular $g^{tt} > 0$. This implies, of course, a certain restriction on 
the coordinate system. In a similar fashion, we use the index $0$ for the 
time component of  a Lorentz vector and greek indices from the beginning 
of the alphabet for the corresponding space components, e.g., $A^a = (A^0, 
A^{\alpha})$, $\alpha = 1,2,3$. (The only exception to those conventions  was 
made in the expressions  (\ref{0}) and (\ref{1}) where  the spacetime 
index of $x^i$ was given values from $0$ to $3$ for simplicity.)  

Let us recall the following theorem for spacelike hypersurface 
integrals \cite{schwinger1}
\begin{equation}\label{4}
0= \int_{\sigma} f_{,i} \de \sigma_k - \int_{\sigma} f_{,k} \de \sigma_i. 
\end{equation}
This results from the fact that the r.h.s. is independent of the choice 
of the hypersurface $\sigma$ (see \cite{schwinger1}), while for the 
specific choice $t= t_0$, it reduces to a two-dimensional surface integral 
which vanishes if an appropriate asymptotical behavior of $f(x)$ is assumed 
(as will always be done throughout this article). From (\ref{4}), we can 
farther deduce the following theorem 
\begin{equation} \label{5} 
\int_{\sigma} f(x) n_i \de \sigma_k (x) = \int_{\sigma} 
f(x) n_k \de \sigma_i(x). 
\end{equation}
Note that in general, $n_i = n_i(x)$, 
but we will omit the argument whenever there is no 
danger of confusion. The above results from $f(x) n_i = f(x) \Phi_{,i}(x) = 
(f(x) \Phi(x))_{,i}- f_{,i} \Phi(x)$, with $\Phi (x) = 0$ on the hypersurface 
we integrate over. Thus, we can apply the previous theorem. 

In particular, we have 
\begin{eqnarray} \label{6}
\int_{\sigma} 
f(x) \delta^i(x-y)  \de \sigma_i &=& 
\int_{\sigma} f(x)  \delta^i (x-y) n^k n_k \de \sigma_i  
\nonumber \\
&=& \int_{\sigma}  f(x) \delta^i (x-y)  n_i n^k \de \sigma_k  
\end{eqnarray} 
which is equal to $f(y)$ if $y$ lies on the hypersurface. (We take the 
convention that all quantities whose arguments are not written explicitely 
are to be taken at the point $x$.) 
Let us introduce 
the following definitions 
\begin{eqnarray} \label{7}
\de \sigma &=& n^i \de \sigma_i \\ \label{8}
\delta_{\sigma} (x-y) &=& \delta^i (x-y) n_i \\ \label{9}
\delta_{\sigma}^i (x-y) &=& \delta_{\sigma}(x-y) n^i 
= \delta^m (x-y) n_m n^i.  
\end{eqnarray}
We can thus write 
\begin{eqnarray} \label{10}
\int_{\sigma}  f(x) \delta^i (x-y) \de \sigma_i (x) &=& \int_{\sigma}  
f(x) \delta^i_{\sigma} (x-y) 
\de \sigma_i(x)  \nonumber  \\ 
&=& \int_{\sigma} f(x)\delta_{\sigma}(x-y) \de \sigma = f(y) 
\end{eqnarray} 
where for the last relation, 
it is assumed that $y$ lies on the hypersurface. Moreover, we have
 $\delta^i (x-y) n_i = \delta_{\sigma}^i (x-y)n_i $. Nevertheless, one should 
not confuse  $\delta^i (x-y)$ which is given explicitely by (\ref{1}), 
with $\delta^i_{\sigma} (x-y)$, which is defined with respect to a specific 
hypersurface. In particular, for $t = t_0$, we have $\delta^i_{\sigma}(x-y) = 
\delta_{t_x t_y}\  \delta^{(3)}(\vec x - \vec y) g^{ti}/g^{tt}$. (The factor 
involving the metric components stems from the normalization $n^2 = 1$, 
which for $t=t_0$ (and thus $n_i = (n_t, 0, 0, 0)$ ) 
leads to $n_t = 1/\sqrt{g^{tt}}$ and $n^i = g^{ti} n_t$.) In particular, 
in flat spacetime, we see that $\delta^i_{\sigma}(x-y)$ has only 
one non-vanishing component, in contrast to (\ref{1}).

For a functional 
\begin{equation} \label{11}
F[\phi,\sigma] = \int_{\sigma}  f[\phi(x)] \de \sigma(x)  
\end{equation}
we define the functional derivative by considering the variation of 
$F$ induced by a variation $\delta \phi$, i.e., if 
\begin{equation} \label{12} 
\delta F = \int_{\sigma} 
 \frac{\delta{f}}{\delta \phi} \delta \phi \ \de \sigma,  
\end{equation}
then we set by definition 
\begin{equation} \label{13}
\frac{\delta F}{\delta \phi(x)} \equiv \frac{\delta{f}}{\delta{\phi}}(x). 
\end{equation}
We are now ready to apply this formalism to classical field theory. 

\section{Covariant Poisson brackets and field momentum} \label{poisson}

Consider a Lagrangian density $\mathcal L = \mathcal L(\phi, \phi_{,m})$. The
corresponding field equations read 
\begin{equation} \label{14}
\pi^m_{\ ,m} = \frac{\partial{\mathcal L}}{\partial \phi}, 
\end{equation}
where 
\begin{equation}\label{15}
 \pi^m \equiv \frac{\partial{\mathcal L}}{\partial \phi_{,m}}  
\end{equation}
will be called  the generalized canonical momentum. Note that $\pi^m $ is 
actually a vector density (i.e., of the form $\sqrt{-g} $ times a vector). 
If there is more than one field, it is understood that there is
 an additional (suppressed) index labeling the different fields 
$\phi^{(A)}$ and the corresponding momentum 
 $\pi^{(A)m}$, and over 
which summation is to be carried out in all our expressions.  
 
According to  our  definition of the functional derivative, we can write 
\begin{equation}\label{16}
\frac{\delta \phi(y)}{\delta \phi(x)} = \delta^i (x-y) n_i, \ \ \ 
\frac{\delta \pi^i(y)}{\delta \pi^m(x)} = \delta^i_m \delta^l (x-y) n_l. 
\end{equation}
Next, we define the classical Poisson brackets by  (see \cite{ozaki}) 
\begin{equation} \label{17}
[A, B]_{\sigma} =  \int_{\sigma} 
\left( \frac{\delta A}{\delta \phi(z)} \frac{\delta B}{\delta \pi^m(z)} - 
\frac{\delta B}{\delta \phi(z)} \frac{\delta A}{\delta \pi^m(z)} 
\right) 
\de \sigma^m(z), 
\end{equation} 
where $\de \sigma^m = g^{mi} \de \sigma_i$. A straightforward 
calculation using  (\ref{16}) as well as the theorem 
(\ref{5}) leads to the following relations
\begin{equation} \label{18}
 [\phi(x), \phi(y)]_{\sigma} = [\pi^m (x), \pi^i(y)]_{\sigma} = 0, \ \ \ 
[\phi(x), \pi^i (y)]_{\sigma} = \delta^i_{\sigma}(x-y)
\end{equation}
The field momentum is defined in the usual way as $\mathcal P_k = 
\int_{\sigma}\sqrt{-g}\  t^i_{\ k} \de \sigma_i$, where $t^i_{\ k} $ is the 
canonical stress-energy tensor. We find  
\begin{equation} \label{19}
\mathcal P_k 
= \int_{\sigma} (\pi^i\phi_{,k} - \delta^i_k \mathcal L )\de \sigma_i. 
\end{equation}
Note that $\mathcal P_k = \mathcal 
P_k( \sigma)$, i.e., $P_k$ depends in general 
on the choice of the hypersurface. In fact, it has been shown in 
\cite{schwinger1} that 
\begin{equation} \label{20}
\frac{\delta }{\delta \sigma} \int_{\sigma} F^i \de \sigma_i = F^i_{\ ,i}, 
\end{equation}
meaning that the functional is independent of the hypersurface whenever  
the integrand is divergence free. (For the precise definition of $ 
\frac{\delta}{\delta \sigma}$, see \cite{schwinger1}.) In our case, this 
means that $\mathcal P_k$ is independent of $\sigma$ if 
$(\sqrt{-g}\ t^i_{\ k})_{,i} = 0$, 
which is the case only in flat spacetime. If we restrict ourselves to 
the  hypersurfaces $t= t_0 = const$, parameterized by $t_0$, this means 
that $\mathcal P_k $ is independent of $t_0$ (i.e., it is conserved) 
whenever $(\sqrt{-g} \ t^i_{\ k})_{,i} = 0$, a well known result.

Next, we show that (for $x \in \sigma$)
\begin{equation} \label{21}
\frac{\delta {\mathcal P_k}} {\delta \mathcal \pi^i(x)} =  \phi_{,k}(x) n_i 
\end{equation}
and 
\begin{equation} \label{22}
\frac{\delta {\mathcal P_k}}
 {\delta \mathcal \phi^(x)} = - \pi^i_{\ ,k}(x) n_i. 
\end{equation}
We show this as follows. Starting from the explicit expression (\ref{19}), 
we find 
\begin{eqnarray*}
\delta \mathcal P_k &=& \int_{\sigma} \left[ \phi_{,k} \delta \pi^i +
\delta \phi_{,k} \pi^i - \delta^i_k \frac{\partial \mathcal L}{\partial \phi}
\delta \phi
- \delta^i_k \frac{\partial \mathcal L}{\partial \phi_{,m}}
\delta \phi_{,m} \right] \de \sigma_i \\
&=& \int_{\sigma} \phi_{,k} \delta \pi^i \de \sigma_i  
- \int_{\sigma} \pi^i_{\ ,k} \delta \phi \de \sigma_i 
+\int_{\sigma} (\delta \phi \pi^i)_{,k} \de \sigma_i 
\\&&
-\int_{\sigma} \left[ \frac{\partial \mathcal L}{\partial \phi}
- (\frac{\partial \mathcal L}{\partial \phi_{,m}})_{,m}\right] \delta \phi 
\de \sigma_k
-\int_{\sigma} (\frac{\partial \mathcal L}{\partial \phi_{,m}} \delta
\phi)_{,m} \de \sigma_k. 
 \end{eqnarray*}
The third integral cancels with the last one in view of the theorem 
(\ref{4}), while the forth integral vanishes on-shell. For the  first 
integral, we write 
\begin{eqnarray*}
\int_{\sigma} \phi_{,k}\delta \pi^i n^m n_m \de \sigma_i 
&=& \int_{\sigma} \phi_{,k}\delta \pi^i n^m n_i \de \sigma_m \\
&=& \int_{\sigma} \phi_{,k}\delta \pi^i n_i \de \sigma, 
\end{eqnarray*} 
where theorem (\ref{5}) has been used in the first step. 
This is now of the form (\ref{12}) and therefore leads to (\ref{21}). 
In the same way, (\ref{22}) follows from the second integral. 

Our expressions are identical in form to those 
derived by  
Schwinger  \cite{schwinger2},  equation (2.91), for the quantized field in 
flat spacetime, where $\pi$ was defined by $\pi^i n_i$ (note that 
Schwinger assumed $n_{i,k} =0 $). 

The Poisson bracket of $\mathcal P_k$ with the canonical field 
variables can now be evaluated and the result is (assuming 
that $x$ lies on  the hypersurface $\sigma$) 
\begin{equation} \label{23}
[ \mathcal P_k, \phi(x) ]_{\sigma} =  - \phi_{,k}(x),  \ \ \ \
[ \mathcal P_k, \pi^i(x) ]_{\sigma} =  - \pi^m_{\ ,k}(x) n_m n^i
\end{equation}
which can be compared to equation (2.92) of \cite{schwinger2}. 

Finally, we point out  that an  alternative approach to the 
definition of the 
Poisson bracket has been presented  in \cite{marsden}. 

\section{Hamilton equations} \label{hamilton}

It is very important to remark that the  equations derived in the previous
section should not be 
confused with the Hamilton equations of motion. The functional  derivatives 
(\ref{21}) and (\ref{22}) are obtained from $\mathcal P_k$ in the form 
(\ref{19}) as it stands. It is in no way understood that $\mathcal P_k$ 
should be 
expressed in terms of $\pi^i $ and $\phi$ only. This cannot be done 
unambiguously anyway, because obviously, (\ref{19}) is not a Legendre 
transformation performed to  replace $\phi_{,i}$ by $\pi^i$, or similar. 
If we write the first term in the form 
$\int \pi^i \phi_{,k} \de \sigma_i = \int \pi^i n_i \phi_{,k} \de 
\sigma$,  we see that $\pi^i$ enters only as $\pi^i n_i$. Thus, at most 
one component of $\phi_{,k}$ could be Legendre transformed in that way. 

The consistent 
way to set up a covariant Hamiltonian theory therefore  consists in Legendre  
transforming the variable $\phi_{,i} n^i$ 
and replace it with $\pi = \pi^i n_i$. 
Such a formalism has been worked out by Ozaki  \cite{ozaki} for the flat space 
case, and we will adopt it here to curved spacetime. Since  
 the formalism is manifestly covariant, there is not much to  
adopt. In fact, the relations of Ozaki should remain identical in form. 
The scope of our presentation is thus basically to check that this is 
indeed the case. Moreover, we will find a slight difference between our 
results and those of Ozaki, which is not directly related to curvature 
effects. 
Following \cite{ozaki}, we define the Hamiltonian by 
\begin{equation}\label{25} 
 h = 
\int_{\sigma} \sqrt{-g}\ t^i_{\ k}  n^k \de \sigma_i 
= \int_{\sigma} (\pi \phi_{,k} n^k
- \mathcal L) \de \sigma,  
\end{equation}
where $\pi$ is defined by 
\begin{equation} \label{26}
\pi = \frac{\partial \mathcal L}{\partial \phi_{,i}} n_i = \pi^i n_i.  
\end{equation}
Note that $h$ should not be confused with the so-called De Donder-Weyl 
Hamiltonian, $\pi^i \phi_{,i} - \mathcal L $, 
 see, e.g., \cite{kanatch}. (See also the so-called polysymplectic 
Hamiltonian field theory \cite{sardan}.) 
The Hamiltonian $h$  thus arises upon Legendre transforming the 
variable $\phi_{,i} n^i $ and replacing  it with $\pi$. 
The Hamiltonian $h$ has first been introduced, in the context 
of special relativistic quantum theory, by Matthews \cite{matthews}.  
For simplicity, 
we consider the case where (\ref{26}) is solvable (in terms of
$\phi_{,m}n^m$). 
Using the formalism of the previous sections, the variation of $h$ leads 
to the following equations 
\begin{eqnarray} \label{27}
\frac{\delta h }{\delta \pi} &=& \phi_{,i} n^i \\ \label{28}
\frac{\delta h }{\delta \phi} &=& - (\pi n^i)_{,i} 
+ \pi^i n^m (n_{i,m} - n_{m,i}).  
\end{eqnarray}
To obtain those relations, theorems (\ref{4}) and (\ref{5}) have to be used 
several times, and in a final step the fact that  
$n^i_{\ ,k} n_i = - n^i n_{i,k}$ (from $n^2 = 1$). 
Equations (\ref{27}) and (\ref{28}) are the Hamilton equations for 
classical field theory in a generally covariant form. They differ 
by the the corresponding equations (11) and (12) of \cite{ozaki} by the 
second term in (\ref{28}), as well as the part $-\pi n^i_{,i}$ 
from the first term, 
which are absent in \cite{ozaki}. In flat 
spacetime, and choosing the hypersurface $t = t_0$, this term vanishes, 
but in curved spacetime, or even in flat spacetime, with a different 
hypersurface, this is not the case anymore. It is not completely clear to 
us where this difference comes from (see however the remarks on the 
conservation of $h$ below).

To get  an idea how this works in practice, consider the (real) scalar field 
Lagrangian 
\begin{equation} \label{29} 
\mathcal L = \sqrt{-g}\ \frac{1}{2} (\phi_{,m} \phi^{,m} - m^2 \phi^2).  
\end{equation}
From (\ref{26}), we find $\pi = \sqrt{-g} \phi_{,m} n^m$. In order to 
express (\ref{25}) in terms of $\pi$, we write 
\begin{equation} \label{30} 
\phi_{,i} =\frac{ \pi}{\sqrt{-g}}\ n_i  + [\phi_{,i} - \phi_{,m}n^m n_i], 
\end{equation} 
which corresponds to a split in normal and tangential components. 
The Hamiltonian is then found in the form 
\begin{equation} \label{31}
h\! =\! 
\int_{\sigma}\! \left[ \frac{1}{2} \frac{\pi^2}{\sqrt{-g}} - \frac{1}{2} 
\sqrt{-g}\ \phi_{,m} \phi^{,m} + \frac{1}{2} \sqrt{\!-g} (\phi_{,m}n^m)^2 
+ \frac{1}{2} m^2 \phi^2 \sqrt{\!-g} \right]\! \de \sigma. 
\end{equation}
Variation and comparison with (\ref{27}) and  (\ref{28}) 
leads to the  Hamilton equations 
\begin{eqnarray} \label{32} 
\frac{\pi}{\sqrt{-g}} = \phi_{,m} n^m  \label{33} \\ 
(\sqrt{-g}\  \phi^{,m})_{,m} + \sqrt{-g}\  m^2 \phi  
-(\sqrt{-g}\  \phi^{,l}n_ln^m)_{,m}  \nonumber  \\ \nonumber 
+ \sqrt{-g}\  \phi^{,i}n^m(n_{i,m}-n_{m,i}) &&
\\ = - (\pi n^i)_{,i} + \pi^i n^m (n_{i,m} - n_{m,i}).   
\end{eqnarray}
As expected, the first equation leads back to the relation between $\pi$ and 
$\phi_{,k}n^k$, and inserting this into the  
 second equation leads to 
the field equation for the scalar field in curved spacetime, 
$(\sqrt{-g}\ \phi^{,m})_{,m}  + \sqrt{-g}\ m^2 \phi = 0$, or simply 
$\Box_g  \phi + m^2 \phi = 0$. Again, the derivation of the second equation 
involves several applications of the theorems (\ref{4}) and (\ref{5}), 
meaning  that the formalism is not really comfortable, even for such simple 
applications. Greater ease could be provided by the definition of  
normal  and tangential  derivatives, 
as has been done in \cite{ozaki}, and observing that certain integrals over 
tangential divergences lead to two dimensional surface integrals 
that can be omitted. This has to be done with care, however, especially in 
curved spacetime.  For instance, 
the following integral over the tangential divergence of a vector field $A^i$, 
$\int_{\sigma} (A^i_{\ ,i} - A^i_{\ ,m} n^m n_i) \de \sigma$ 
leads to $\int_{\sigma} (A^i n^k_{\ ,k} - A^k n^i_{\ ,k}) \de \sigma_i$, 
which is not always  zero. 

It is instructive to see the explicit expression of $h$ for the hypersurface 
$t = t_0$, i.e., $n_i = (1/{\sqrt{g^{tt}},0,0,0})$, and 
$n^i = g^{ik} n_k = g^{it}/ \sqrt{g^{tt}}$, such that 
$\de \sigma = n^i \de \sigma_i = \sqrt{g^{tt}}\  \de^ 3 x$. 
The scalar field Hamiltonian (\ref{31}) then takes the form (recall that 
spacetime indices are denoted  $m = (t, \mu)$, $\mu = 1,2,3$, where it is 
assumed that $t$ is timelike, i.e., $g^{tt} > 0$) 
\begin{eqnarray} \label{34}
h &=&  \int_{\sigma} \left[ 
\frac{1}{2}  \frac{\sqrt{g^{tt}}}{\sqrt{-g}} \ \pi^2
- \frac{1}{2} 
\sqrt{-g g^{tt}}\ \phi_{,m} \phi^{,m}  \right. \nonumber \\
&& \ \ \ \    \left.
+ \frac{1}{2} \sqrt{-g g^{tt}}\ (\phi_{,t} \sqrt{g^{tt}}  + \phi_{,\mu} 
\frac{g^{t\mu}}{\sqrt{g^{tt}}})^2 
+ \frac{1}{2} m^2 \phi^2 \sqrt{-g g^{tt}} \right]\de^3 x. 
\end{eqnarray}
The momentum in terms of $\phi_{,m}$ is of the form 
\begin{equation} \label{35}
\pi = \sqrt{-g}\ \phi_{,m}n^m = \sqrt{-g}\ \left(\phi_{,t} \sqrt{g^{tt}} + 
\phi_{,\mu} \frac{g^{\mu t} }{\sqrt{g^{tt}}}\right). 
\end{equation}
Note that, since $t$ is timelike, the three dimensional 
tensor  $g^{\mu\nu}$ is negative definite, and therefore $h$ is positive.  
In flat spacetime, $\pi$ reduces to $\phi_{,t}$, and the Hamiltonian 
takes the conventional form $h = \int_{\sigma} \frac{1}{2} (\pi^2 - 
\phi_{,\mu}\phi^{,\mu} + m^2 \phi^2) \de^3 x$. It is hard to imagine how 
the Hamiltonian (\ref{34}) could have been guessed without having at our 
disposal a manifestly covariant formalism. 

The above expressions take a simpler form if we use the Arnowitt-Deser-Misner 
(ADM) 
decomposition of the metric tensor, given by $g_{tt} = N^2 - \tilde g_{\mu\nu}
N^{\mu\nu}$, $g_{t\mu} = -N_{\mu}$ and $g_{\mu\nu} = - \tilde g_{\mu\nu}$. 
Then we have (for the same hypersurface as above) $\pi = \sqrt{\tilde g}\ (
\phi_{,t} - \phi_{,\mu} N^{\mu})$, and 
\begin{displaymath}
h = \int_{\sigma} \left[ \frac{1}{2} \frac{\pi^2}{\sqrt{\tilde g}} + 
\frac{1}{2} \sqrt{\tilde g}\ \tilde g^{\mu\nu} \phi_{,\mu} \phi_{,\nu} 
+ \frac{1}{2} \sqrt{\tilde g}\ m^2 \phi^2 \right] \de^3 x,  
\end{displaymath}
where three dimensional indices are raised and lowered with $\tilde 
g_{\mu\nu}$ and its inverse $\tilde g^{\mu\nu}$

Finally, let us derive the Poisson bracket of $h$ from (\ref{31}) 
with the canonical 
variables $\phi$ and $\pi$. For this, we need an expression for  
$\delta h/ \delta \pi^i$. Consider some  functional  
$F = \int_{\sigma} f \de \sigma $. According to  (\ref{12}), the variation 
with respect to $\pi$ is given by 
\begin{displaymath}
\delta F = \int_{\sigma} \frac{\delta f}{\delta \pi} \delta \pi \de \sigma 
 = \int_{\sigma} \frac{\delta f}{\delta \pi} (\delta \pi^i n_i) 
\de \sigma 
= 
\int_{\sigma} (\frac{\delta f}{\delta \pi} n_i) \delta \pi^i \de \sigma
\equiv 
 \int_{\sigma} \frac{\delta f}{\delta \pi^i} \delta \pi^i \de \sigma
\end{displaymath}
meaning that we have 
\begin{equation} \label{36}
 \frac{\delta F}{\delta \pi^i} = \frac{\delta F} {\delta \pi} n_i. 
\end{equation}
We now easily find the following relations 
\begin{equation} \label{37}
[h, \phi(x)]_{\sigma} \!=  - \phi_{,i}(x) n^i, \ \ [h, \pi(x)]_{\sigma} \!= 
- (\pi(x) n^i)_{,i} + \pi^i n^m (n_{i,m}\! -\! n_{m,i}). 
\end{equation}
Again, the second relation differs by the terms $\pi n^i_{,i}$ and 
$\pi^i n^m(n_{i,m}-n_{m,i})$ from the 
result of \cite{ozaki}. The first relation suggests to interpret  
$h$ as the generator of translations along $n^i$, i.e., normal to the 
spacelike hypersurface. 
We will therefore occasionally refer to $h$ as evolution operator.  
In particular, for $t = t_0$ and assuming a flat 
spacetime, $h$ becomes 
 equivalent to the time component of $\mathcal P_i$, and 
thus to the generator of time translations. It is important, however, 
that the direct interpretation of $h$ as evolution operator is only valid 
for expressions that do only depend on the field $\phi$. For expressions 
involving $\pi$, or $n^i$ and $g_{ik}$, the  evolution is not 
directly given in terms of $h$. One could eventually modify the Hamiltonian 
in order to get a relation of the form $[\tilde h, \pi] 
= - \pi_{,m} n^m$, but this 
would not really solve the problem for general expressions. Obviously, 
any Hamiltonian   will satisfy 
$[\tilde h, n_i] = [\tilde h ,g_{ik}] = 0$ as long as the 
metric is treated as a background field, and therefore, it is not possible 
to construct an evolution operator satisfying 
$[\tilde h,f] = - f_{,i} n^i$ for a 
general expression $f(\phi, \pi, n^i, g_{ik})$. 

To illustrate what this means in practice, consider the case of the free 
electromagnetic field. We have $\pi^{ik} = \frac{\partial \mathcal L}
{\partial A_{i,k}}$, which is antisymmetric, 
and thus, we have the primary constraints 
$\Phi(x)= \pi^{ik} n_in_k = 0$. For consistency of the Hamiltonian theory, 
we have to require that $\Phi(x)$ remains zero during the evolution of the 
system, i.e., $\delta \Phi= \Phi(x^i + \epsilon n^i) - \Phi(x^i) 
= \epsilon \Phi_{,i} n^i = 0 $ (where $\epsilon$ is an infinitesimal
parameter), which leads to secondary constraints. Therefore, the secondary 
constraints are not simply obtained by requiring  $[ h, \Phi] = 0$. 
Instead, in our case, we have $\Phi(x) = \pi^{ik} n_in_k = \pi^i n_i $, 
and thus 
$\delta \Phi = (\pi^i_{\ ,k} 
n^k) n_i + \pi^i (n_{i,k} n^k)$, where $\pi^i_{\ ,k} n^k$ is 
evaluated from  equation (\ref{37}) in the form $\pi^i_{\ ,k} n^k = 
- [h, \pi^i]_{\sigma} - \pi^i n^l_{,l} + \pi^{ik}n^m(n_{k,m}- n_{m,k})$. 
 We have checked that this leads indeed 
to the secondary constraint $\pi^{ik}_{\ \ ,k} n_i = 
(\sqrt{-g}\ F^{ik})_{,k} n_i = 0$, which is 
the covariant expression for Gauss' law. Any other expression, e.g., 
$(\sqrt{-g} F^{ik} n_i)_{,k}$ or similar would not be in accordance with 
the field equations. For the sake of completeness, we give the 
explicit form of the Maxwell Hamiltonian  
\begin{displaymath} 
h =\! \int_{\sigma} 
\left[ -\frac{1}{2} \frac{\pi^i \pi_i}{\sqrt{-g}} + \pi^i A_{m,i} n^m
+ \frac{1}{4} \sqrt{-g}\ F^{lm}F_{lm}+ \frac{1}{2} \sqrt{-g}\ F_{li}n^i 
F^{ml}n_m \right] \de \sigma.  
\end{displaymath} 
with $\pi^i = \pi^{ik} n_k = \sqrt{-g} F^{ik} n_k$. The velocities have been 
eliminated using 
$A_{i,k} = - \frac{\pi^i}{\sqrt{-g}} n_k + (A_{i,k} - F_{li}n^l n_k)$.  
The relations corresponding to (\ref{37}) for  $A_i$ and $\pi^i$ are easily 
established. They are identical to the scalar field case (just replace 
$\phi$ with $A_i$,  $\pi$ with $\pi^i$ and $\pi^{k}$ with $\pi^{ik}$). 
It is now an easy task to derive the secondary constraints. 

Finally, we choose the hypersurface $t = const$ and use 
the ADM parameterization. The Hamiltonian then simplifies to  
\begin{displaymath}
h = \int_{\sigma} \left[\frac{1}{2} \frac{\pi^{\mu} \pi_{\mu}}{N^2
    \sqrt{\tilde g}} + \pi^{\mu} A_{t,\mu} N^{-2} - 
\pi^{\mu} A_{\nu,\mu} N^{\nu} N^{-2} + \sqrt{\tilde g}\  
\frac{1}{4} F^{\mu\nu} F_{\mu \nu} \right ] \de^3 x,  
\end{displaymath}
where indices are raised and lowered with $\tilde g_{\mu\nu}$. 
For the commutation relations, we remind that $\delta_{\sigma}(x-y) = 
\delta^i(x-y) n_i = \delta^{3}(x-y) N$, and therefore we have 
$[A_i(x), \pi^k(y)]_{\sigma} = \delta^k_i N \delta^{(3)}(x-y)$ at equal
times. The  conventional choice  for the momentum, in the framework of 
the ADM formalism, differs by a factor $N$ from ours, namely $\tilde \pi^i = 
\frac{\partial \mathcal L}{\partial A_{i,t}}$, instead of 
$\pi^i = \frac{\partial \mathcal L}{\partial A_{i,k}}n_k$, resulting therefore 
in a simpler  commutation relation, 
$[A_i(x), \tilde \pi^k(y)]_{\sigma} = \delta^k_i  \delta^{(3)}(x-y)$.  
(In other words, we work with $\pi^i = 
\pi^{ik} n_k$, while the conventional choice is $\pi^i = \pi^{it}$.) 
This is merely a matter of convenience. 
The conventional choice would appear  unnatural when written in an 
explicitely covariant form. 

It has also 
been claimed in \cite{ozaki} that (in flat spacetime), $h$ is independent 
of the choice of the hypersurface $\sigma$. We do not agree with that 
statement. 
According to Schwinger \cite{schwinger1}, an integral $\int_{\sigma} f \de 
\sigma_i $ is independent of $\sigma$ whenever we have $f_{,i} = 0$. 
For $h = \int_{\sigma} t^i_{\ k} n^k \de \sigma_i $
 this would be the case for $(t^{ik} n_k)_{,i} = 0$, or, in view of the 
special relativistic conservation law for the canonical stress-energy tensor, 
for $t^{ik} n_{i,k} = 0$. For a general $t^{ik}$, this can only be the case 
for $n_{i,k} = 0$, which holds only for specific hypersurfaces. 

Similar, in curved spacetime, $h$ is independent of $\sigma$ for 
$(\sqrt{-g}\ t^{ik} n_k )_{,i} = 0$. 
In the framework of general relativity, 
we have shown \cite{leclerc} 
that the canonical stress-energy tensor $t^i_{\ k}$ is 
equivalent to the metric (Hilbert) tensor $T^{ik}$, i.e., they differ by 
a relocalization term and  lead both 
to the same momentum vector $\mathcal P_k$. 
However, in general, $t^{ik}$ does not 
satisfy $t^{ik}_{\ \ ;k} = 0$ and moreover, it is easy to show that 
$t^i_{\ k}$ and $T^{ik}$ 
do not lead to the same $h$. (Note that the canonical tensor is not 
a genuine tensor, and should be referred to as  pseudo-tensor. However, in 
order to avoid confusion with the pseudo-stress-energy tensor of the 
gravitational field itself, we will stick to the expression canonical 
tensor, which is widely  used in literature.) 
In the specific 
case of  the scalar field, however, $t^{ik}$ incidentally coincides 
with the Hilbert tensor, and therefore satisfies the general relativity 
relation $t^{ik}_{\ \ ;i} = 0$, or ($\sqrt{-g}\ t^{ik})_{,i} + \sqrt{-g}\ 
\hat \Gamma^k_{li}\ t^{il} = 0 $, where $\hat \Gamma^k_{li}$ is the 
Christoffel connection. It is then straightforward to show that  
$(\sqrt{-g}\ t^{ik} n_k )_{,i} = 0$ (i.e., that $h$ is independent of
$\sigma$), whenever we have $t^{ik} (n_{i;k} + n_{k;i}) = 0$, 
where we made use of the symmetry of $t^{ik}$. For this to hold 
in general, we must  
 have $n_{i;k} + n_{k;i} = 0$. We thus must have a timelike 
Killing vector, orthogonal to a spacelike  hypersurface, meaning that 
spacetime must be static. (In general, even if such a Killing vector 
$\xi^i$ exists,
it will not be normalized. In order to find a conserved quantity, one 
will have to replace $h$ by $\int_{\sigma} \sqrt{-g}\ 
t^i_{\ k} \xi^k \de \sigma_i$.) 
Therefore, only for static spacetimes can we choose $n_i$ (or rather $\xi_i$) 
in a way that $h$ is 
conserved. (Recall that this result is limited to the case where 
$t^{ik}_{\ \ ;i} = 0$.) In no way, however, is $h$ generally independent of 
the choice of $\sigma$. 

In flat spacetime, for the hypersurface $t= t_0$, we have 
$h = \mathcal P_t$, which justifies the conventional interpretation of 
$\mathcal P_t$ as field Hamiltonian in special relativistic field theory. 
Nevertheless, they are fundamentally different quantities from a theoretical 
standpoint, 
and according to the above considerations, 
only $h$ should be referred to 
with the name Hamiltonian, since it is the quantity that, when expressed 
in terms of $\phi$ and $\pi$, leads upon variation 
to the canonical Hamilton equations.

\section{Second quantization} \label{second}

One of the major successes of the canonical Hamiltonian formulation 
of mechanics and field theory is its direct relation to quantum theory 
by means of a suitable correspondence principle. 
On the other hand, a special relativistic theory can unambiguously be written 
in a manifestly generally covariant form, if we exclude direct curvature
couplings. In the presence of bosonic 
fields, this results in the replacement of the non-dynamical Minkowski metric 
by a spacetime dependent metric tensor $g_{ik}$. According to the principle of 
equivalence, this tensor is interpreted as dynamical gravitational field. 
In the presence of spinor fields, we need a tetrad field $e^a_i$ ($a,b,\dots$ 
denote Lorentz vector indices) and the metric arises as derived  quantity 
$g_{ik}= e^a_i e^b_k \eta_{ab}$. The theory is invariant under local 
Lorentz gauge transformations 
(in addition to general coordinate transformations)
$e^a_i \rightarrow \Lambda^a_{\ b}(x) e^b_i$, and 
 the special relativistic limit is given by $e^a_i = \delta^a_i$, i.e., 
when we have identification of the tangent Lorentz space with the spacetime 
manifold. Obviously, in that limit, the residual symmetry is the global 
Lorentz group, acting at the same time in tangent space and in spacetime, 
because in order to have 
$\delta^a_i = \tilde e^a_i(\tilde x)  
= \Lambda^a_{\ b} \ e^b_k (x)\ \frac{\partial x^k}{\partial 
\tilde x^i}$, where $e^a_i = \delta^a_i$, $\Lambda^a_{\ b}$ must be the 
inverse of $  \frac{\partial x^k}{\partial \tilde x^i}$, meaning in 
particular that 
the coordinate transformation $  \frac{\partial x^k}{\partial \tilde x^i}$
has to be a Lorentz transformation. 
     The important thing 
is that there is again only one covariant form to a given special relativistic 
theory. (Note, however, that this one to one correspondence 
does not hold for gravitational theories which 
involve additional variables, like the (independent) Lorentz connection 
in Poincar\'e gauge theory. In such a case additional tensors arise (e.g., 
torsion), that can be coupled at will to the matter fields, without destroying 
the general covariance. It turns out, however, that 
for the canonical quantization process, only the metric (or tetrad) structure 
is of importance.)

In summary, assuming the ideal case,  
to each classical special relativistic theory corresponds 
exactly one  second quantized theory (correspondence principle), and to each 
special relativistic theory corresponds exactly one (eventually 
up to additional, non-metric gravitational tensor fields) 
generally covariant theory. It is therefore clear that, if there exists 
a manifestly covariant second quantized theory, then it can only be obtained 
by applying the correspondence principle to the special relativistic 
theory written in  generally  
covariant form. If this does not lead to a consistent theory, this means 
that the quantization process cannot be done in a covariant, and 
thus ultimately, in an observer independent way. One would then  
 have to fix first the coordinate system, and perform the quantization 
afterwards.  It is clear that this would ultimately  mean that the 
principle of equivalence does not hold on a quantum level. 
Here, we assume that the second quantization can be performed 
 in a manifestly  covariant way. Taking into account 
gravitational effects (from the background curvature) is then, at least from 
a formal point of view,  a trivial issue.

Our starting point is the Poisson bracket (\ref{17}), and  
second quantization is achieved by replacing the field $\phi$ 
and the canonical momentum $\pi^i$ by operators acting in a Hilbert space 
and satisfying 
the relations (\ref{18}), where the Poisson bracket of the 
classical fields is replaced by the commutator of the corresponding operators, 
\begin{equation} \label{38}  
[A, B]_{\sigma}  \rightarrow  \frac{1}{i} [A, B] = \frac{1}{i} (A B - BA). 
\end{equation}
Explicitely, the canonical commutation  relations (CCR's) read 
\begin{equation} \label{39} 
 i[\phi(x), \phi(y)] = i[\pi^m (x), \pi^i(y)] = 0, \ \ \ 
i [\pi^i(x), \phi(y)] =  \delta^i_{\sigma}(x-y),  
\end{equation}
where $x$ and $y$ are on the hypersurface $\sigma$. In particular, 
$x-y$ is thus spacelike. (Recall that 
the  commutation relations between 
fields at timelike separations can only be obtained with the help of the 
field equations for the specific theory under investigation (propagator), 
and  explicit expressions are only available for free fields (interaction 
representation), see \cite{schwinger1}. The direct application of 
Schwinger's formalism to curved spacetime can only be achieved if we 
expand the metric around a flat background, $g_{ik} = \eta_{ik} + h_{ik}$. 
Not only does this mean giving up general covariance, but moreover, 
in view of the inverse of $g_{ik}$ occurring in standard matter Lagrangians, 
the explicit expressions in terms of $h_{ik}$ contain infinite 
series already at this early stage of the theory.) The above CCR's are 
identical to those used by Schwinger for the special relativistic theory 
\cite{schwinger1} and are equivalent to those  used in 
 \cite{nikolic} on a  curved  background.  

For the specific hypersurface $t =t_0$, we find from (\ref{1}) and (\ref{9}) 
that the last relation reduces to 
\begin{equation} \label{40} 
i [\pi^t (x), \phi (y)] =  \delta^{(3)}(\vec x - \vec y) \delta_{t_x t_y}, 
\end{equation} 
and $i [\pi^{\mu}(x), \phi(y)] = 0$. Since both points are assumed to lie 
on $\sigma$ anyway, 
we can omit the factor $\delta_{t_x t_y}$, and we find 
the conventional equal time 
commutation relation at $t_0$ between the field and its canonical momentum. 
We see that, although (\ref{39}) involves four components $\pi^i$, there 
is only one component that plays the role of the canonically conjugate field, 
namely the component normal to the hypersurface, $\pi  = \pi^i n_i$. 
In terms of $\pi$, we have the following  CCR's 
\begin{equation} \label{41} 
i [\phi(x), \phi(y)] = i [\pi (x), \pi(y)] = 0,  \ \ \  
i [\pi(x), \phi(y)]  = 
 \delta_{\sigma} (x-y),   
\end{equation} 
which is the form used in \cite{nikolic}. 

Let us apply this to the scalar field Lagrangian (\ref{29}). The field 
momentum (\ref{19}) is given by 
\begin{eqnarray} \label{42} 
\mathcal P_k 
&=&  \int_{\sigma} (\pi^i\phi_{,k} - \delta^i_k \mathcal L )\de \sigma_i
\nonumber \\ &=&  \int_{\sigma} \pi^i\phi_{,k} \de \sigma_i  - \frac{1}{2} 
\int_{\sigma} 
 \sqrt{-g}\ (\phi_{,m} \phi^{,m} - m^2 \phi^2) \de \sigma_k,  
\end{eqnarray}
where $\pi^i = \sqrt{-g} \phi^{,i} $. It is important to recall that 
$\mathcal P_k$ has nothing to do with a Hamiltonian. Although 
$\mathcal P_t$ is conventionally referred to  
as field Hamiltonian, as we have seen in the previous section, $\mathcal P_k$ 
 is not a Legendre transformation and is not a quantity which is used to 
determine equations of motion. In particular, it is not understood that 
$\phi_{,m}$ should be replaced by $\pi^m$ in (\ref{42}). From the form 
(\ref{42}) and the CCR's, 
we can deduce the commutation relations of $\mathcal P_k $ with $\phi$ and 
$\pi^i$. Note that, for the evaluation of $[\mathcal P_k,\pi^i(x)]$, 
we have to write, e.g., for the first term from (\ref{42}) 
$\int \pi^l(y) [\phi_{,k}(y), \pi^i(x)]\de \sigma_l(y)
= i \int \pi^l(y) (\delta^i_{\sigma}(y-x))_{,k} \de \sigma_l(y) 
= - i \int \pi^l_{\ ,k}(y)\delta^i_{\sigma}(y-x)\de \sigma_l(y)  
+ i \int(\pi^l(y) \delta^i_{\sigma}(y-x))_{,k} \de \sigma_l(y)$  
 (derivatives acting on $y$), where the second term cannot simply be 
omitted. (To our knowledge, there is no generalization of the rule 
$\delta'(x-y) f(x)= - f'(x) \delta(x-y)$ for our $\delta$ functions.) 
Instead, it cancels with an opposite  term coming from the second  
term  in (\ref{42}), as can be shown using theorem (\ref{4}) and 
the definition of $\pi^m$. The final relations are found in the form  
\begin{equation} \label{43} 
i [P_k, \phi(x)] =  \phi_{,k}(x),  \ \ \  i [P_k, \pi^i(x)] 
= \pi^m_{\ ,k}(x) n_m n^i, 
\end{equation}
which are in direct correspondence to (\ref{23}). It is understood that 
$x$ lies on the same hypersurface $\sigma$ (with normal vector $n_i$) 
that  has been used in the definition of $\mathcal P_k$. Note that 
(\ref{43}) are on-shell relations. In terms of $\pi = \pi^i n_i$, 
the above relations read 
\begin{equation} \label{44} 
i [P_k, \phi(x)] =  \phi_{,k}(x),  \ \ \  i [P_k, \pi(x)] 
= \pi^m_{\ ,k}(x) n_m, 
\end{equation}
From its action on the field $\phi$, we can identify, 
as expected, $\mathcal P_k$ with the  generator of spacetime translations. 
The same equations have been derived (assuming flat spacetime and 
$n_{i,k} = 0$) by Schwinger \cite{schwinger2}, see equation (2.92).

On the other hand, for the evolution operator $h$ (\ref{31}), we 
find 
\begin{eqnarray} 
i[h,\phi] &=&\frac{\pi}{\sqrt{-g}}, \nonumber \\  
i[h, ,\pi] &=&  
-(\sqrt{-g}\  \phi^{,m})_{,m} - \sqrt{-g}\  m^2 \phi  \nonumber 
\\&& +(\sqrt{-g}\  \phi^{,l}n_ln^m)_{,m}  
- \sqrt{-g}\  \phi^{,i}n^m(n_{i,m}-n_{m,i}). \label{45}
\end{eqnarray} 
If we assume the correspondence principle and write the  equations (\ref{37})
in the form 
\begin{equation} \label{46}
i [h, \phi(x)] =  \phi_{,i}(x) n^i,  \ \ \ \ 
i [h, \pi(x)] =  
 (\pi(x) n^i)_{,i} - \pi^i n^m(n_{i,m}-n_{m,i}),  
\end{equation}
then (\ref{45}) leads consistently to the field equations. Those equations 
have been derived for the first time in \cite{matthews} for flat spacetime
 (and $n_{i,k} = 0$). We see that, thanks to the manifestly covariant 
formalism, the generalization to curved spacetime does not present any 
difficulties, at least on a formal level.

\section{Dirac field} \label{dirac}

\subsection{Classical theory}

The Dirac equation in presence of  gravitational fields (see e.g.,
\cite{obukhov}) has been 
studied in our previous article \cite{leclerc1} 
with the focus on the relation between 
the Dirac Hamiltonian and the time component of the field momentum. 
Here, we will extend this analysis to the case of the quantized 
theory.

Recall the Dirac Lagrangian 
\begin{equation} \label{47}
\mathcal L = e \left[ \frac{i}{2}(\bar \psi \gamma^i \De_i \psi 
- \De_i \bar \psi \gamma^i 
 \psi) - m \bar \psi \psi \right],  
\end{equation}
with $e = \det e^a_i = \sqrt{-g}\ $, 
$\De_i \psi = \partial_i \psi - \frac{i}{4} \Gamma^{ab}_{\ \ i} 
\sigma_{ab}\psi $ and $\gamma^i = e^i_a \gamma^a$, where $\gamma^a$ 
are the constant Dirac matrices and $\sigma_{ab}$ the Lorentz 
generators. 
Recall that  we use latin indices from the begin of the alphabet
($a,b,c \dots $) for 
tangent space quantities and from the middle ($i,j,k \dots $)  
for curved spacetime quantities. 
The  Lorentz  connection
$\Gamma^{ab}_{\ \ i}$ ($\Gamma^{ab}_{\ \ i} = - \Gamma^{ba}_{\ \ i}$)
and the tetrad  $e^a_i$  
transform under an (infinitesimal)  local Lorentz transformation  
 $\Lambda^a_{\ b} = \delta^a_b + \epsilon^a_{\ b}$ according to  
$ \delta \Gamma^{ab}_{\ \ i}  = - 
 \partial_i \epsilon^{ab}
- \Gamma^a_{\ ci}  \epsilon^{cb} - \Gamma^b_{\ ci}  \epsilon^{ac}, 
 \  \delta e^a_i =   \epsilon^a_{\ b} e^b_i$.  
The Lagrangian (\ref{47}) is invariant under such transformations 
if $\psi $ undergoes the gauge transformation  
$ \psi \rightarrow \exp{[(-i/4) \epsilon^{ab} \sigma_{ab}}]\psi $.
The Dirac equation is derived in the form 
\begin{equation} \label{48}
i \gamma^i \nabla_i \psi = m \psi,   
\end{equation}
where $ \nabla_i \psi = \partial_i \psi - \frac{i}{4} \tilde 
\Gamma^{ab}_{\ \ i}\sigma_{ab} \psi $. 
The connection $\tilde \Gamma^{ab}_{\ \ i}$ is the connection that 
effectively couples to the spinor field. If we work in the framework 
of general relativity, then $\Gamma^{ab}_{\ \ i} = \Gamma^{ab}_{\ \
  i}(e)$, i.e., the connection is not an independent field and can 
be expressed in terms of the tetrad and its derivatives. In that case, 
$\tilde \Gamma^{ab}_{\ \ i} = \Gamma^{ab}_{\ \ i} $ (and $\De_i = \nabla_i$). 
If the connection is considered to be an independent field (Poincar\'e gauge 
theory), then we have $\Gamma^{ab}_{\ \ i} = \Gamma^{ab}_{\ \ i}(e) 
+ K^{ab}_{\ \ i}$, where $K^{ab}_{\ \ i}$ is the contortion tensor. 
However, only the totally antisymmetric 
part of $K^{ab}_{\ \ i}$ remains in the field equations. 
Therefore, the effective connection is given by 
$\tilde \Gamma^{ab}_{\ \ i} = \Gamma^{ab}_{\ \ i} + \tilde K^{ab}_{\ \ i}$, 
where $\tilde K^{ab}_{\ \ i} = e_{ci} K^{[abc]}$.  Just as was the case in 
\cite{leclerc1}, those  
differences 
between general relativity and Poincar\'e gauge theory are not  
related to our specific  discussion, and  the  
use of the symbols 
$\nabla_i$ and $ \tilde \Gamma^{ab}_{\ \ i} $ is a convenient  way of 
treating both cases at the same time. 

In Schroedinger 
form, the Dirac equation can be written as 
\begin{equation}\label{49}
H \psi = i \partial_t \psi  
\end{equation}
Using again greek indices $\alpha, \beta \dots $ 
to denote the spatial part of $a,b \dots $ (i.e., $ a = (0,\alpha)$), 
and indices $\mu, \nu \dots $ for the spatial part of the 
spacetime indices $i,j,k \dots $ and $t $ for the time component, 
(e.g., $m = (t, \mu)$), we find 
\begin{equation} \label{50}
H = \frac{1}{g^{tt}} ( \gamma^t m - i \gamma^t \gamma^{\mu} \nabla_{\mu}) 
- \frac{1}{4} \tilde \Gamma^{ab}_{\ \ t} \sigma_{ab}. 
\end{equation}
In the flat limit ($\tilde \Gamma^{ab}_{\ \ i} = 0,\  e^a_i  = \delta^a_i$), 
this reduces to the well known expression $H = \gamma^0 m 
- i \gamma^0 \gamma^{\mu} \partial_{\mu} = \beta m + \vec \alpha \cdot \vec p$.

Further, we define a manifestly covariant inner product in 
Dirac space  
\begin{equation}\label{51}
(\psi_1, \psi_2) = \int_{\sigma}\sqrt{-g}\ 
\bar \psi_1 \gamma^i \psi_2 \de \sigma_i, 
\end{equation} 
where the integration is performed over a spacelike hypersurface. We have 
shown in \cite{leclerc1} that $H$ is not in general 
hermitian with respect to this 
scalar product. Thus, by the (Dirac space) operator identity 
$H = i \partial_t$, the time evolution operator $i \partial_t$ is not 
hermitian. 
A similar situation holds for the translation operators $i \partial_{\mu}$, 
and altogether, it was shown that the following  operator ($m = 0,1,2,3$) 
\begin{equation} \label{52} 
 \tilde p_m  =  i \partial_m + \frac{i}{2} \partial_m \ln\sqrt{-g g^{tt}},  
\end{equation}
is hermitian. (Whenever expressions are used that 
are not manifestly covariant, as is the case with (\ref{52}), 
it is understood that they refer to the hypersurface $t = t_0$, which has 
been used exclusively in \cite{leclerc1}. The reason is that an explicit 
$3+1$ split has already been performed by writing down equation (\ref{49}). 
This can be avoided by writing $H_{\sigma} = i n^m \partial_m \psi $ instead, 
and then repeat the steps performed in \cite{leclerc1}. The fact remains 
that the operator $i \partial_m $ is not hermitian. Moreover, in 
order to find the explicit form (\ref{52}), it is assumed that 
$e^t_{\alpha} = 0$ for $\alpha = 1,2,3$, which can always be achieved 
by a suitable Lorentz rotation, since $e^t_a $ is a timelike Lorentz 
vector.)   
Moreover, it was shown that the  expectation values of (\ref{52}) are given by 
the field momentum, i.e., 
\begin{equation} \label{53} 
\mathcal P_k = (\psi, \tilde p_k \psi), 
\end{equation}
where 
\begin{equation} \label{53b} 
\mathcal P_k 
= \int_{\sigma}\sqrt{-g}\  t^i_{\ k} \de \sigma_i = 
\int_{\sigma}  \left[\frac{\partial \mathcal L}
{\partial (\partial_i\psi) } \ \partial_k \psi 
\ +\  \partial_k \bar \psi\ \frac{\partial \mathcal L}
{\partial (\partial_i \bar \psi)} \ -\  \delta^i_k \mathcal L\right]    \de
\sigma_i. 
\end{equation}
In particular, if $\mathcal P_k$ is to play the role of the generator of 
spacetime translations in the quantum theory, then it seems strange that 
this operator does not coincide with the expectation value of the  
momentum  $i \partial_k$ of the corresponding classical theory, as is the 
case in flat space (recall, e.g., the relation 
$\mathcal P_t = \mathcal H = (\psi,H \psi) 
= \int \psi^{\dagger} H \psi \de^3 x$ of the special relativistic theory).   
Therefore, it is of interest to investigate the situation in the second 
quantized theory. 
Surprisingly, it will turn out that, in  contrast to 
the case of the boson field (see equation (\ref{43})), the field momentum 
$\mathcal P_k$ does not coincide with the generators of spacetime translations 
in curved spacetime. 

\subsection{Fermion Poisson Bracket and second quantization} \label{hermitian} 

The adaptation  of the Poisson bracket   (\ref{17}) to the 
case of spinor fields is straightforward. Since we expect again 
a correspondence principle to lead to  second quantization, and since 
fermions are quantized with anticommutators, the Poisson bracket 
for fermions, $\{A, B\}_{\sigma}$ should be symmetric in $A$ and $B$. 
Therefore, we simply replace the minus sign with a plus sign in (\ref{17}) 
and, from a formal point of view,  we are done. (The cases for fermion 
and boson brackets can also be formally treated at the same time, see, e.g., 
\cite{ozaki}, where you can also find the generalized Jacobi identity 
for such brackets.) The larger problem concerns the correct choice of 
the canonical variables. Formally, we have in (\ref{47})  two independent
 fields $\psi$ and $\bar \psi$, related by hermitian conjugation, giving 
us the two corresponding momentum variables $\pi^i = \frac{i}{2}\sqrt{-g}\ 
\bar \psi \gamma^i $ and $\bar \pi^i = -\frac{i}{2}    \sqrt{-g}\ \gamma^i 
\psi$ (recall that only the component in $n_i $ direction is the physical  
canonical momentum). Obviously, the variables $(\psi, \bar \psi, \pi, 
\bar\pi)$ are not independent, and we have two constraints in the theory, 
which are easily shown to be second class. In order to 
get a consistent theory with such constraints, it is necessary to modify 
the classical Poisson bracket (in order to exclude non-physical degrees 
of freedom) 
before passing over to the second quantization (see Dirac \cite{dirac} for 
details). Fortunately, since the quantization of the special relativistic 
Dirac theory is well known, we can directly inspire ourselves from the 
corresponding theory and put it into a generally covariant form. 
More details on the problems arising from 
the second class constraints will be given in section \ref{strictly}.  

The consistent way is to use $(\psi, \pi)$ as canonically conjugate variables, 
with $\pi^i = i \sqrt{-g}\ \bar \psi \gamma^i $, and thus to define 
the following Poisson bracket
\begin{equation} \label{54}
\{A, B\}_{\sigma} =  \int_{\sigma} 
\left( \frac{\delta A}{\delta \psi^M (z)} \frac{\delta B}{\delta \pi_M^m(z)} + 
\frac{\delta A}{\delta \pi_M^m(z)} \frac{\delta B}{\delta \psi^M(z)} \right) 
\de \sigma^m(z), 
\end{equation} 
where we use capital letters $K,L,M,N\dots$ to denote spinor
indices. More precisely, upper indices are used for spinors 
transforming like $\psi$ 
under Lorentz gauge transformations, $\psi \rightarrow \Lambda \psi$, 
while lower indices are used for quantities transforming with the inverse, 
like $\bar \psi\rightarrow \bar \psi \Lambda^{-1}$. 
Thus, e.g., the Dirac matrices transform as $\gamma^a 
\rightarrow \Lambda \gamma^b \Lambda^{-1} \Lambda^a_{\ b} \equiv \gamma^a$, 
where $\Lambda^a_{\ b} $ is the corresponding Lorentz transformation in the 
vector representation. Thus, we write $(\gamma^a)^M_{\ N}$. Further, 
if we denote with an upper 
dotted index a spinor transforming like $\psi^{\dagger} 
\rightarrow \psi^{\dagger} \Lambda^{\dagger}$, and with a dotted lower index  
spinors transforming with the inverse $(\Lambda^{\dagger})^{-1}$, 
then the spinor 
metric $\gamma^0$ (used in $\bar \psi = 
\psi^{\dagger} \gamma^0$) takes two lower indices, a dotted and an undotted, 
$(\gamma^0)_{\dot M N}$ and 
is invariant under $(\Lambda^{\dagger})^{-1} \gamma^0 \Lambda^{-1}$. 
Note that $(\gamma^0)_{\dot M N}$,  
although numerically identical with the zeroth component of $\gamma^a$ 
 is not a component of a Lorentz vector. In summary, the situation is 
identical to the special relativistic case, only that the 
Lorentz coordinate transformations are replaced by local Lorentz gauge 
transformations (not acting on the argument of $\psi(x)$). 

The following Poisson brackets are easily derived (for $x,y$ on $\sigma$) 
\begin{eqnarray} \label{55}  
\{\psi^N(x), \psi^M(y)\}_{\sigma} 
= 0, \ \ \{\pi^i_N (x), \pi^k_M(y)\}_{\sigma} = 0, \nonumber \\  
\{\psi^N(x), \pi^i_M(y)\}_{\sigma} = \delta^N_M \delta^i_{\sigma}(x-y). 
\end{eqnarray}
Those relations are identical to those used (in the quantum theory) by 
Schwinger (\cite{schwinger1}). Note that the asymmetry between $\psi $ and 
$\bar \psi$ is only apparent. The corresponding relation for $\bar \psi$ and 
$\bar \pi = \gamma^0 \pi^{\dagger} $ is easily found by hermitian 
conjugation of (\ref{55}). In other words, we could equally well 
start with the canonical pair $(\bar \psi, \bar \pi)$ and construct 
the Poisson bracket accordingly.

Let us note that $\pi^i = i \sqrt{-g}\ \bar \psi \gamma^i $ can be 
inverted to $\bar \psi = - i (\sqrt{-g})^{-1} \pi^in_i \gamma^k n_k$, 
where the relation $\gamma^in_i \gamma^k n_k = 1 $ has been used. 

For the momentum vector $\mathcal P_k = \int_{\sigma} \sqrt{-g}\ t^i_{\ k}
\de \sigma_i$, we find 
\begin{equation} \label{56} 
\mathcal P_k = \int_{\sigma} \frac{i}{2} \sqrt{-g}\ \bar \psi \gamma^i
\psi_{,k} \de \sigma_i - \int_{\sigma}
\frac{i}{2} \sqrt{-g}\ \bar \psi_{,k} \gamma^i 
\psi \de \sigma_i,  
\end{equation} 
where we have used the fact that $\mathcal L = 0 $ on shell. (Indeed, 
it is not hard to show that we can write $\mathcal L = \frac{1}{2}(\bar \psi 
\frac{\delta \mathcal L}{\delta \bar \psi} + \frac{\delta \mathcal L}{\delta 
\psi} \psi)$, where the  variations $\delta \mathcal L/\delta \psi$  
and 
$\delta \mathcal L/\delta \bar \psi $ 
vanish in virtue of the field equations. Thus, on the set of solutions 
of the field equations, $\mathcal L$ is zero.) 
Next, we perform a partial integration of the second term, use the  
expression for $\bar \psi $ in terms of $\pi^i$, as well as the 
fact that $\int_{\sigma} (\sqrt{-g}\ \bar\psi \gamma^i \psi)_{,k} \de \sigma_i
= 0$, as is shown by using (\ref{4}) and the on-shell relation 
$(\sqrt{-g}\ \bar\psi \gamma^i \psi)_{,i} = 0$ (see \cite{leclerc1}). The 
result is 
\begin{equation} \label{57} 
\mathcal P_k = \int_{\sigma} \pi^i \psi_{,k} \de \sigma_i + \frac{1}{2} 
\int_{\sigma} \frac{1}{\sqrt{-g}} \pi^k n_k \gamma^l n_l 
(\sqrt{-g} \gamma^i)_{,k}  \psi \de \sigma_i. 
\end{equation}
Performing another partial integration, we can also write 
\begin{equation} \label{58} 
\mathcal P_k = - \int_{\sigma} \pi^i_{\ ,k} 
\psi \de \sigma_i + \frac{1}{2} 
\int_{\sigma} \frac{1}{\sqrt{-g}} \pi^k n_k \gamma^l n_l 
(\sqrt{-g} \gamma^i)_{,k}  \psi \de \sigma_i. 
\end{equation}
Those expressions can be used to evaluate $\delta \mathcal P_k/ \delta \pi^i$ 
and $\delta \mathcal P_k/ \delta \psi$ respectively, which are needed 
to find the Poisson brackets between $\mathcal P_k$ and the 
canonical field variables. The result is 
\begin{eqnarray} \label{59}
[ \mathcal  P_k, \psi ]_{\sigma}   &=& - \psi_{,k}  - \frac{1}{2} 
\frac{1}{\sqrt{-g}} \ \gamma^m n_m (\sqrt{-g}\ \gamma^i)_{,k} n_i \psi 
\\ \label{60}
{[} \mathcal  P_k, \pi^l {]}_{\sigma}   &=& - \pi^i_{,k} n_i n^l + \frac{1}{2} 
\pi^k n_k (\gamma^m n_m) (\sqrt{-g}\ \gamma^i)_{,k} n_i n^l.  
\end{eqnarray}
Note that at the left hand side, the Poisson bracket (\ref{17}) has 
been used. This is because $\mathcal P_k$ is bilinear in the spinor 
fields and as such, of bosonic nature. 
Comparing these results with (\ref{23}), we see that they both differ in 
the appearance of an additional term. We will discuss this term 
shortly. For the moment, let us just remark that it vanishes for 
flat spacetime. 

We now proceed to  second quantization. We use again the correspondence 
principle 
\begin{equation} \label{61}  
\{A, B \}_{\sigma}  \rightarrow  \frac{1}{i} \{A, B\} = \frac{1}{i} 
(A B + BA),  
\end{equation}
leading to the canonical anticommutation relations 
\begin{eqnarray} \label{62}
i \{\psi^N(x), \psi^M(y)\} = 0, \ \ i \{\pi^i_N (x), \pi^k_M(y)\} = 0,
\nonumber \\  
i \{\psi^N(x), \pi^i_M(y)\} = - \delta^N_M \delta^i_{\sigma}(x-y), 
\end{eqnarray}
and from the expressions (\ref{57}) and (\ref{58}), we derive 
\begin{eqnarray} \label{63}
[ \mathcal  P_k, \psi ]  &=& - i  \psi_{,k}  - \frac{i}{2} 
\frac{1}{\sqrt{-g}} \ \gamma^m n_m (\sqrt{-g}\ \gamma^i)_{,k} n_i \psi
\\ \label{64}
{[} \mathcal  P_k, \pi^l {]}   &=& - i \pi^i_{,k} n_i n^l + \frac{i}{2} 
\pi^k n_k (\gamma^m n_m) (\sqrt{-g}\ \gamma^i)_{,k} n_i n^l,   
\end{eqnarray}
where commutators are used at the left hand side. Again, those relations 
differ from  their bosonic counterparts (\ref{44}) 
by the additional term at the right hand side. 

Quite generally, for an operator $\mathcal O$ 
that corresponds to the expectation value 
of a Dirac space operator $O$, i.e, 
$\mathcal O = (\psi, O \psi) = 
- i \int_{\sigma} \pi^i O \psi \de \sigma_i $ (see (\ref{51})), we obtain  
\begin{equation} \label{65}
[\mathcal O, \psi] = -  O \psi. 
\end{equation}
In particular, for $\mathcal P_k $, we should thus have, according to 
 (\ref{53}) 
\begin{equation}
[\mathcal P_k, \psi] = -  \tilde  p_k \psi, 
\end{equation}
where $\tilde p_k$ is the hermitian momentum operator. 
Indeed, for the hypersurface $t= t_0$ and assuming a gauge $e^t_{\alpha} = 0$, 
$\alpha = 1,2,3$, we find from (\ref{63})
\begin{equation} \label{66} 
[\mathcal P_k, \psi] 
=  - i\left( \partial_k + \frac{1}{2} (\partial_k \ln\sqrt{-g g^{tt}}\ 
)\right) 
\psi,  
\end{equation}
which  is in perfect agreement with (\ref{52}).  

Similar relations are obtained for the  
 evolution operator $h = \int_{\sigma}  
\sqrt{-g}\  t^i_{\ k} n^k \de \sigma_i $ defined in (\ref{25}). 
The same additional term  will appear in $[h,\psi]$ and $[h, \pi]$, both in 
the classical and  in the quantum case. The analysis is straightforward and 
does not lead to new insight.

We conclude that $\mathcal P_k$ is not the generator of spacetime 
translations. (Neither is $h$ the generator of translations along 
$n_i$.) This is in direct 
correspondence  with our result of \cite{leclerc1}, namely 
that $\mathcal P_k$ is not the expectation value of the operator 
$p_k = i \partial_k$, but rather of the hermitian operator $\tilde  p_k$. 
Also, from (\ref{65}) it is clear what operator corresponds to the 
generators of translations: It is the expectation value of $p_k$. 
Thus, if we define $\mathcal P^{(1)}_k = (\psi, p_k \psi)$, then we have 
$[\mathcal P^{(1)}_k, \psi ] = -  p_k \psi  
= - i \partial_k \psi$. There is only one 
problem: the operator $p_k$ is not hermitian. Therefore, there is 
a second operator $\mathcal 
P_k^{(2)} = (p_k \psi, \psi)$, which could equally well 
be used to define the expectation value of $p_k$. How can we decide 
which of those operators, if any,  corresponds to the physical field momentum? 
The reason for those ambiguities will become clear in the next section.

Finally, let us remark that the results are not an artifact of the 
quantization process. The same deviation from the bosonic case  
has been  obtained in the classical case, namely equations 
(\ref{59}) and (\ref{60}). The reason is also obvious: It is 
the result of the fact that, due to the constraints in the theory,  
we were forced to deviate slightly from the canonical procedure. Thus, 
ultimately, the problem originates from the fact that the Lagrangian theory 
contains two independent field variables $\psi$ and $\bar \psi$, but 
the corresponding Hamiltonian theory is constructed only from one canonically 
conjugate pair of variables, $\psi $ and $\pi^i$. This is not avoidable 
(anything else leads to inconsistencies), but it is also not unproblematic
 and leads to problems such as the above,  concerning the interpretation of 
the field momentum. This will become very clear in the next section. 

Let us also indicate that we can construct a Lorentz generator in the 
form 
\begin{displaymath} 
\Sigma_{ab} = - \frac{i}{4} \int_{\sigma} \pi^i \sigma_{ab} \psi \de \sigma_i, 
\end{displaymath}
with $\sigma_{ab} = \frac{i}{2} [\gamma_a, \gamma_b]$, which generates 
a Lorentz gauge transformation on the field $\psi$, i.e., we have 
\begin{displaymath}
[\Sigma_{ab}, \psi]_{\sigma} = - \frac{i}{4} \sigma_{ab} \psi, 
\end{displaymath}
in consistence with the invariance of (\ref{47}) under $\delta \psi = 
- \frac{i}{4} \epsilon^{ab} \sigma_{ab}\psi$ with 
arbitrary (antisymmetric) gauge parameters 
$\epsilon^{ab} = \epsilon^{ab}(x)$. 

Obviously, the same construction of $\Sigma_{ab}$ can be used in the 
quantum theory. Note that in contrast to  the special relativistic 
theory, the above generator does not contain the orbital term $\sim \int  
\pi(x^{k} \partial^{i} - x^{i} \partial^{k})\psi \de^3x$. This 
is because in generally covariant theories, the spinor field $\psi$ 
transforms us a scalar under coordinate transformations, its spinor nature 
being exclusively related to the internal Lorentz transformations. 
In other words, the above operator generates gauge transformations, and not 
Lorentz coordinate transformations. In special relativity, both 
transformations are directly related, since the spinor transformation 
acts simultaneously on the argument of $\psi$, , i.e., $\psi'(x') = 
S \psi(x)$, while in our case, 
we have $\psi'(x) = S(x)\psi(x) $, which, although a local transformation, 
acts at a given point $x$. 
(As is well known, the special relativistic spinor (Lorentz) transformations 
of $\psi$ cannot be generalized to the  full diffeomorphism  group, 
and therefore the only way is to treat $\psi$ as an invariant under 
those transformations, and to replace the spinor transformations by 
Lorentz gauge transformations, unrelated to coordinate transformations.)
 Note, however, that $\Sigma_{ab}$ is not 
a conserved quantity. A covariant conservation law for the 
angular momentum, based on the 
invariance of $\mathcal L$ under Lorentz gauge transformations, 
has been given  in \cite{leclerc}. The orbital part is thereby expressed 
in terms of the Hilbert stress-energy tensor. This does still not lead, 
in the absence of gravitational fields, to the special relativistic
expression. In fact, in order to obtain a generalization of the special 
relativistic conservation law, one will have to restrict the coordinate 
transformation group to the Lorentz group, such that, for  the remaining 
class of coordinate systems, one can establish a one-to-one correspondence 
between internal Lorentz gauge transformations and  
the residual Lorentz coordinate transformations. 
Then, one can construct a complete set of 10 Poincar\'e generators, see,  
e.g., \cite{nakanishi}. Such a construction, based on an explicit  
symmetry breaking,  is not needed 
for our purpose which focuses on the momentum operator, and we will 
 continue to treat the coordinate and gauge transformations separately.

\subsection{Relocalization of field momentum} \label{reloc}

Consider the Dirac Lagrangian (\ref{47}) in the following form 
\begin{equation} \label{67}
\mathcal L = \sqrt{-g} \left[ \frac{i}{2}(\bar \psi \gamma^i \partial_i \psi 
- \partial_i \bar \psi \gamma^i \psi) \right] + \mathcal L_{int},  
\end{equation}
where $\mathcal L_{int}$ contains the interaction terms between the 
spin connection and the spinor fields, as well as the mass term. 
This can equivalently be written in the form 
\begin{equation} \label{68} 
\mathcal L = \sqrt{-g}\ i \bar \psi \gamma^i \partial_i \psi 
+ \frac{i}{2} \bar \psi (\gamma^i \sqrt{-g})_{,i} \psi 
- (\sqrt{-g}\ \bar \psi \gamma^i \psi)_{,i}   + \mathcal L_{int},  
\end{equation}
Omitting the surface  term, we find the following Lagrangian
\begin{equation} \label{69} 
\mathcal L^{(1)} = \sqrt{-g}\ i \bar \psi \gamma^i \partial_i \psi 
+ \frac{i}{2} \bar \psi (\gamma^i \sqrt{-g})_{,i} \psi 
  + \mathcal L_{int},  
\end{equation}
which leads upon variation with respect to $\psi$ and $\bar \psi$ to 
the same  equations as (\ref{67}). Note that the role of the field 
variable $\bar \psi$ has essentially been  reduced to that of a Lagrange 
multiplier. Further, we recall that $\mathcal L$ is on-shell zero  
(see remark after (\ref{56})) 
 and that $(\sqrt{-g}\ \psi \gamma^i \psi)_{,i}$ is 
also zero  (charge conservation, see, e.g., \cite{leclerc1}). As a result, 
$\mathcal L^{(1)}$ too is on-shell zero. The canonical stress-energy tensor 
\begin{equation}\label{70}
\sqrt{-g}\ t^{(1)i}_{\ \ \ k} =  \frac{\partial \mathcal L^{(1)}}
{\partial (\partial_i\psi) } \ \partial_k \psi 
\ +\  \partial_k \bar \psi\ \frac{\partial \mathcal L^{(1)}}
{\partial (\partial_i \bar \psi)} \ -\  \delta^i_k \mathcal L^{(1)} 
\end{equation}
therefore reduces to (note that $\mathcal L_{int}$ does not contain 
derivatives and therefore does not contribute in the first two terms of 
(\ref{70}))  
\begin{equation} \label{71} 
\sqrt{-g}\ t^{(1)i}_{\ \ \ k} = \sqrt{-g}\  \bar \psi \gamma^i \psi_{,k}, 
\end{equation}
and the corresponding field momentum is of the form  
\begin{equation} \label{72}
\mathcal P_k^{(1)} = i \int_{\sigma}\sqrt{-g}\ \bar \psi \gamma^i \psi_{,k} 
\de \sigma_i, 
\end{equation}
which differs from (\ref{57}) by the second term in the latter. The fact 
that the omission of a surface term in $\mathcal L$ leads to a modification 
of the stress-energy tensor (a so-called relocalization) is well known. 
That it leads to a modification of the integrated momentum is a little bit 
more surprising. It is, however, quite natural. The reason can be traced back 
to the fact that a strict separation between the energy and momentum of 
the dynamical fields (in our case, the Dirac field) on one hand and 
the non-dynamical background fields (in our case, gravity) is devoid of
physical sense. In other words, it is rather a matter of convention which 
amount of energy (momentum) is attributed to one or the other part, only 
the total energy (momentum) being  of physical relevance. 

We wish to point out that this is not a particularity of gravity. Instead of 
a general discussion, simply consider the following (albeit unrealistic) 
surface Lagrangian (in flat spacetime) 
\begin{displaymath} 
\mathcal L_{surf} = B_{i,k} {}^*\!F^{ik}, 
\end{displaymath} 
where $B_i$ is a dynamical vector field, and ${}^*\!F^{ik}$ is the dual of 
the Maxwell tensor $F_{ik} = A_{k,i} - A_{i,k}$ for a given electrodynamic 
background field $A_i$. Since ${}^*\!F^{ik}_{\ \ ,k}= 0 $ identically, 
the above is indeed a total divergence. The stress-energy tensor for the 
dynamical field $B_i$ takes an additional term   
$\tilde t^i_{\ k} 
= \frac{\partial \mathcal L_{surf}}{\partial B_{l,i}}B_{l,k} - 
\delta^i_k \mathcal L_{surf}$ satisfying $\tilde 
t^i_{\ k,i} = {}^*\! F^{lm}_{\ \ ,k} 
B_{l,m}$, which is not identically zero, and thus the addition of $\mathcal 
L_{surf}$ (to the Lagrangian for the dynamical field $B_i$) 
does not merely lead to a relocalization of  $t^i_{\ k}$, 
but rather to a modification of the integrated momentum. Exactly the 
opposite  term in $\mathcal P_k$ will be induced in the corresponding  
momentum vector of the field $A_{m}$, as is easily shown, such that 
the surface term does indeed not contribute to the total momentum.   

Thus, although usually, surface terms lead to a relocalization of the 
stress-energy, but leave the integrated momentum vector unchanged, 
in the presence of background fields, this is not true anymore. Apart from 
a relocalization, a momentum transfer between dynamical and background 
fields is induced by surface terms. Both relocalizations and  momentum 
transfers, 
however, should not be physically relevant. It is quite 
 a matter of convention whether we attribute, e.g., 
 the potential energy of an electron in the Coulomb field of a proton 
either to the electron or to the electromagnetic field. 

Having reduced $\bar \psi$ to a Lagrange multiplier, we consider $\psi$ as 
the only true field variable in $\mathcal L^{(1)}$ and define 
\begin{equation}\label{73}
\pi^i = \frac{\partial\mathcal L^{(1)}}{\partial \psi_{,i}} = i \sqrt{-g}\ 
\bar \psi   \gamma^i, 
\end{equation}
and postulate again the anticommutation relations 
\begin{eqnarray} \label{74}
i \{\psi^N(x), \psi^M(y)\} = 0, \ \ i \{\pi^i_N (x), \pi^k_M(y)\} = 0,
\nonumber \\ 
i \{\psi^N(x), \pi^i_M(y)\} = - \delta^N_M \delta^i_{\sigma}(x-y).   
\end{eqnarray}
According to (\ref{72}), we have 
 $\mathcal P^{(1)}_k = \int_{\sigma} \pi^i \psi_{,k} \de \sigma_i$, and  
the following commutators are straightforwardly evaluated 
\begin{eqnarray} \label{75}
[ \mathcal  P^{(1)}_k, \psi ]   &=& - i  \psi_{,k} = - p_k \psi \\ \label{76}
{[} \mathcal  P^{(1)}_k, \pi^l {]}   &=& - i \pi^i_{,k} n_i n^l,   
\end{eqnarray}
which are now in complete correspondence to the bosonic case (\ref{43}) and 
$\mathcal P^{(1)}_k $ can be interpreted as generator of spacetime
translations. Consistently, we also have that $P^{(1)}_k = (\psi, p_k \psi)$.  

It is needless to say that instead of $\bar \psi$, we can also eliminate 
$\psi$ as dynamical variable and write 
\begin{equation} \label{77} 
\mathcal L^{(2)} = - \sqrt{-g}\ i \partial_i \bar \psi \gamma^i \psi 
- \frac{i}{2} \bar \psi (\gamma^i \sqrt{-g})_{,i} \psi 
  + \mathcal L_{int},  
\end{equation}
where the same surface term with the opposite sign has been 
omitted this time. The 
canonical momentum is now given by $\bar \pi^i  
= - i \sqrt{-g}\ \gamma^i \psi$, 
and assuming 
 $i \{\bar\psi_N(x), \bar \pi^{iM}(y)\} = + \delta^M_N
\delta^i_{\sigma}(x-y)$ (for  
 consistency with (\ref{74}), we have 
to  use the opposite sign, such that one relation results from hermitian 
conjugation of the other) 
we find
\begin{equation} \label{79}
[ \mathcal  P^{(2)}_k, \bar \psi]   =  - i  \bar \psi_{,k}, 
\ \ \ [ \mathcal  P^{(2)}_k, \bar \pi^l ]   =  - 
i  \bar \pi^i_{\ ,k} n_i n^l,     
\end{equation} 
where we have $\mathcal P^{(2)}_k = (p_k \psi, \psi)$. 
  
\subsection{The strictly canonical way} \label{strictly}

For completeness, we will briefly outline the problems that arise 
from the second class constraints if one tries to  follow strictly the 
canonical procedure. The issue is not related to the 
covariant formalism, neither 
to the presence of the gravitational background fields and should be known 
to most readers from the corresponding special relativistic theory. 

Formally, starting from the Dirac Lagrangian (\ref{47}), 
one is let to define $\pi^i = \partial \mathcal L/ \partial \psi_{,i} 
= \sqrt{-g}\ 
(i/2) \bar \psi \gamma^i $ and $\bar \pi^i 
= \partial L/ \partial \bar \psi_{,i}
= - \sqrt{-g}\ (i/2) \gamma^i \psi $, and according to the general procedure, 
one assumes the following anticommutation relations 
\begin{eqnarray*} 
i \{\psi^N(x), \psi^M(y)\} = 0, \ \ i \{\pi^i_N (x), \pi^k_M(y)\} = 0,
\\ i \{\psi^N(x), \pi^i_M(y)\} = - \delta^N_M \delta^i_{\sigma}(x-y), \ \
i \{\bar \psi_N(x), \bar \psi_M(y)\} = 0, \\ i \{\bar \pi^{iN} (x), \bar 
\pi^{kM}(y)\} = 0,\ \ 
i \{\bar \psi_N(x), \bar \pi^{iM}(y)\} = + \delta^M_N \delta^i_{\sigma}(x-y), 
\end{eqnarray*}
where again, the sign difference is necessary in order for consistency, since 
 hermitian conjugation  of the relations in the first line  leads 
to those in the  second. There is also 
a classical realization for those relations, based on  the 
Poisson bracket 
\begin{eqnarray*}  
\{A, B\}_{\sigma} &=&  \int_{\sigma} 
\left( 
\frac{\delta A}{\delta \psi^M (z)} \frac{\delta B}{\delta \pi_M^m(z)}  
-\frac{\delta B}{\delta \bar 
\pi^{mM}(z)}  \frac{\delta A}{\delta \bar \psi_M (z)}\right. \\
&&\left. +\frac{\delta A}{\delta \pi_M^m(z)} \frac{\delta B}{\delta \psi^M(z)} 
-\frac{\delta B}{\delta \bar 
\psi_M(z)} \frac{\delta A}{\delta \bar \pi^{mM}(z)} 
\right) 
\de \sigma^m(z).  
\end{eqnarray*} 
Further, is it understood that $\{\bar \psi_N, \psi^M \} 
= \{\bar \pi^{iM}, \pi^k_N\}
= \{\bar \psi_N, \pi^i_M\} = \{\bar \pi^{iN}, \psi^M\}=0$, 
since they are anticommutators between different components of the canonical 
field variables $\Psi^A = (\psi, \bar \psi)$ and momentum variables   
$\Pi^i_A =
(\pi^i, \bar \pi^i)$, where $A = 1,2\dots 8$. The momentum vector 
$\mathcal P^{(can)}_k = \int_{\sigma} \sqrt{-g}\ t^i_{\ k} \de \sigma_i$ takes 
the simple form (since $\mathcal L = 0$) 
\begin{displaymath} 
\mathcal P^{(can)}_k  = \int_{\sigma} (\pi^i \psi_{,k} 
+ \bar \psi_{,k} \bar \pi^i ) \de \sigma_i
\end{displaymath}
which leads immediately to 
\begin{displaymath}
[ \mathcal  P^{(can)}_k, \psi ]  = - i  \psi_{,k}, \ \ \ 
[ \mathcal  P^{(can)}_k, \bar \psi ]  = - i  \bar \psi_{,k},  
\end{displaymath}
as was to be expected, since we faithfully sticked to the canonical 
procedure. 

Unfortunately, the above procedure is not correct. The reason, as we 
have pointed out, is found in the constraints of the theory. Indeed, 
from the expressions for $\pi^i$ and $\bar \pi^i$, we find two 
constraints 
\begin{displaymath} 
\Phi_1^i = \pi^i - \sqrt{-g}\ \frac{i}{2} \bar \psi \gamma^i \equiv 0, \ \ \ 
\Phi_2^i = \bar \pi^i + \sqrt{-g}\ \frac{i}{2} \gamma^i  \psi \equiv 0.    
\end{displaymath}
Note that we use here a manifestly covariant form of 
Dirac's analysis \cite{dirac} with a four component momentum $\pi^i$. 
Actually, the physical constraints are found by considering  
the constraints along $n_i$, i.e., $\Phi_{\alpha}= 
\Phi_{\alpha}^i n_i = 0$, $\alpha = 1,2$, 
 involving only  the true canonical momentum $\pi^i n_i$. 
For our purposes, the component form 
can equally well be used, but one should have in mind that there are two 
(not eight) physical constraints (per point).  

Using the anticommutation relations between the fields, we can evaluate 
\begin{displaymath} 
\{\Phi_1^i(x), \Phi_2^k(y) \} = -\frac{1}{2}\sqrt{-g}\ 
(\delta^i_{\sigma}(x-y) \gamma^k + \delta_{\sigma}^k(x-y) \gamma^i),  
\end{displaymath}
or, contracting with $n_i n_k$, $\{\Phi_1(x), \Phi_2(y)\} = -
\sqrt{-g}\ \delta_{\sigma}(x-y) \gamma^l n_l\neq 0$. To be precise, the 
above relations, according to Dirac \cite{dirac} have to be evaluated 
with the classical Poisson bracket, because we can proceed to the second 
quantization only after having dealt with the constraints. The result is 
of course the same. Such constraints, which do not possess  a vanishing 
Poisson bracket, are referred to as second class constraints. In contrast 
to first class constraints (with vanishing Poisson brackets between each
other), which are relatively easily dealt with (first class 
constraints are the kind that arises in gauge theories and  are 
imposed on the  physical states of the theory $\Phi | \Psi>_{phys} = 0$, 
recall, e.g., the Gupta-Bleuler quantization), second class constraints 
are more difficult to handle, and have, in some way, to be eliminated 
from the theory. According to Dirac, a necessary step, before 
the correspondence principle can be applied, is to modify the Poisson 
bracket. If this is not done, we have obviously an inconsistent theory 
(since $\{\Phi_1, \Phi_2\} \neq 0 $ for $\Phi_1 = \Phi_2 = 0$). 

The inconsistency of the theory can also be seen directly, 
noting that, e.g.,  the anticommutation relation $\{\psi(x), \pi^i(y)\} = 
i \delta^i_{\sigma}(x-y) $ is not consistent with $\{ \psi, \bar \psi\} = 0$, 
as can be seen by inserting $\pi^i = \sqrt{-g}\ (i/2) \bar \psi \gamma^i $ 
into the first relation. Also, for instance, 
the bracket $\{\mathcal P^{(can)}_k, \psi\}$ 
gives a different result (namely zero) 
if we replace, e.g., $\pi^i $ by $\bar \psi$ in 
the expression for  $\mathcal P^{(can)}_k$. Thus, quite obviously, we have 
too many canonical variables in our formalism.  
A general method of how one deals with 
second class constraints, removing the non-physical degrees of freedom from 
the theory, 
has been outlined in \cite{dirac}. The crucial step is to replace the 
Poisson bracket $\{A,B\}_{\sigma}$ by a modified bracket $\{A,B\}^*_{\sigma}
= \{A,B\}_{\sigma}- \{A,\Phi_{(m)}\}_{\sigma}c^{mn} 
\{\Phi_{(n)}, B\}_{\sigma}$, 
where $c^{mn} = \{\Phi_{(m)}, \Phi_{(n)}\}_{\sigma}^{-1}$ is the inverse of 
the  
matrix formed from the second class constraints $\Phi_{(m)}$. Second
quantization  is achieved upon application of  the correspondence principle  
on this modified Poisson bracket. 
This method, when applied  to the Dirac field,  leads  to the  
theories 
described in the previous sections, \ref{hermitian} and 
\ref{reloc}, depending on whether one starts from the manifestly hermitian 
Lagrangian, or from the surface term modified Lagrangian,  respectively. 

Here, instead,  we were led to the same theories 
  inspiring  ourselves directly from the quantization of the 
special relativistic theory, since, we repeat, the problems described in
this section are completely unrelated to the covariant formalism and to the 
eventual presence of gravitational fields.  

\subsection{Discussion}

We have given two approaches to the  canonical quantization for the Dirac 
field in curved spacetime. The first way, presented in section
\ref{hermitian}, is based on the manifestly hermitian Lagrangian and preserves 
the symmetry between the fields $\psi$ and $\bar \psi$. The field momentum 
$\mathcal P_k$ is equal to the expectation value of the hermitian 
momentum $\tilde p_k$, but does not  generate  spacetime
translations on the quantum fields. The second way, presented in 
section \ref{reloc}, starts 
from a non-hermitian Lagrangian, where one of the field variables  
$\psi$ or $\bar \psi$ has been eliminated by adding a suitable surface term to 
the Lagrangian. In this approach, the field momentum is equal to the 
expectation value of the  non-hermitian momentum operator $p_k = i \partial_k$ 
and corresponds to the generator of the spacetime translations. 

On a classical level, we have traced back the differences between both 
approaches to a momentum transfer from the Dirac field to the gravitational 
background,  induced by the surface term in the Lagrangian. Since the 
assignment of the potential (or interaction) energy to one or 
another of the interacting fields (Dirac and gravitational) is rather a matter 
of convention, both approaches should be physically equivalent. In order to 
establish this equivalence on the quantum level, further investigations 
are necessary. We suspect that both approaches can be related by a change 
of  representation in operator space, 
involving a non-unitary transformation, very similar to 
the Dirac space transformation that has been presented in \cite{leclerc1} 
to relate the hermitian momentum $\tilde p_k$ to the non-hermitian 
translation operator $p_k = i \partial_k$. 

To illustrate this, we consider the manifestly hermitian theory of section 
\ref{hermitian}, and choose again, for simplicity,  the hypersurface 
$t=t_0$ and the gauge $e^t_{\alpha} = 0$, such that equation  (\ref{66}),  
\begin{displaymath}  
[\mathcal P_k, \psi] 
=  - i\left( \partial_k + \frac{1}{2} (\partial_k \ln\sqrt{-g g^{tt}}\ 
)\right) 
\psi,  
\end{displaymath}
is valid. Next, we perform a transformation $A$ in Hilbert space, acting on 
states $|\Phi>$ and on operators $\mathcal O$ 
\begin{displaymath}
|\Phi> \rightarrow A|\Phi>, \ \ \ \mathcal O \rightarrow \tilde \mathcal O = 
A \mathcal O A^{-1}, 
\end{displaymath}  
where 
\begin{displaymath} 
A = (-g g^{tt})^{-\frac{1}{4}}.
\end{displaymath}
This specific operator is diagonal in both Hilbert and Dirac space, and 
thus commutes, e.g., with $\psi$, $\pi^i$, $\psi_{,k}$ etc., as well as 
with $\mathcal P_k$ from (\ref{57}). Therefore, we find $\tilde \psi = 
A \psi A^{-1} = \psi$, and similarly, $\tilde \pi^i = \pi^i$,  
$\tilde \mathcal P_k = \mathcal P_k$, and so on. On the other hand, 
for $p_k = i \partial_k$, we find 
\begin{displaymath}
\tilde \partial_k = A  \partial_k A^{-1} =  \partial_k + \frac{1}{2}  
(\partial_k \ln\sqrt{-g g^{tt}}\ ),  
\end{displaymath}
leading to the hermitian operator $\tilde p_k $ from (\ref{52}). 
Therefore, in the new representation, we can write 
\begin{displaymath}  
[\tilde \mathcal P_k, \tilde \psi] = [\mathcal P_k, \psi] 
=  
- i\left( \partial_k + \frac{1}{2} (\partial_k \ln\sqrt{-g g^{tt}}\ )\right) 
\psi    
=
- i  \tilde \partial_k  \tilde \psi,   
\end{displaymath}
showing that $\mathcal P_k$ can indeed be interpreted as the generator of  
translations, but in a different representation. Those manipulations 
seem to indicate that there is indeed a direct connection between the 
hermitian formulation of the theory and the approach  of section \ref{reloc}, 
where $\mathcal P^{(1)}_k$ appears directly as generator of translations. 

One can object that, if both approaches are related by 
a non-unitary transformation, they are still not physically equivalent, 
since such a transformation will not leave invariant, e.g.,  the expectation 
values of physical operators, and this will ultimately jeopardize the 
probability interpretation of the theory. This, however, is not necessarily 
the case, if the scalar product in Hilbert space is transformed simultaneously 
with the change of representation. In fact, on a quantum mechanical level, 
we have already shown in \cite{leclerc1} how this works explicitely. 
Namely, we have to use the explicitely covariant scalar product 
 $(\psi_1, \psi_2) = \int \sqrt{-g}\  
\bar \psi_1 \gamma^i \psi_2 \de \sigma_i$ in the hermitian case, 
while after the non-unitary transformation, we simply use the 
\textit{flat} scalar product 
$<\psi_1,\psi_2>= \int \psi_1^{\dagger} \psi_2 \de^3 x$. 
In this way,  we find the same expectation values before and after the
non-unitary transformation that relates $p_i$ to $\tilde p_i$. 
In a sense, as we have outlined in \cite{leclerc1}, 
the non-unitarity of the transformation is absorbed by the change 
of the scalar product. 

One can thus expect that a similar 
construction is possible on a field theoretical level, namely 
that both representations are equivalent if simultaneously with the 
non-unitary transformation, an appropriate change of the scalar product 
in Hilbert space is performed. (Technically, this means actually the 
introduction of a new Hilbert space.) Before one can 
conclude any  further, one will first have to construct an explicit 
representation that satisfies the  anticommutation relations (\ref{62}). 
(This is not a trivial matter, since for anticommuting fields, 
we can not use the conventional representation of the bosonic  case, 
where the field $\phi$ is a  multiplication 
operator and $\pi$ a functional derivative 
operator $- i  \frac{ \delta }{\delta \phi}$. In the fermion case, one will 
have to use Grassman variables to construct a representation.) 

Thus, both the so-called {\it manifestly hermitian} theory and the 
{\it non-manifestly hermitian} theory can turn out to be  hermitian. 
The term {\it hermitian} is actually meaningless as long as we do not 
specify a Hilbert space (and in particular a scalar product) with respect to 
which the theory should be hermitian. (All we can say for now is that 
the manifestly hermitian theory is based on a (classically) {\it real} 
Lagrangian, but there is no direct correspondence between {\it real}
and {\it hermitian}, except if restrict ourselves to specific scalar 
products. In this article, nevertheless, we have followed the 
usual convention to call real Lagrangians hermitian.)

Finally, it is interesting to remark that in flat spacetime, 
both approaches are  
equivalent, and the generator of translations in either 
approach is automatically given in terms of the field momentum.  
 It is actually quite common, in standard textbooks, to 
use the non-hermitian version of the Dirac Lagrangian and to omit the  
discussion related to the second class constraints completely. 
Our analysis shows that the situation in curved spacetime needs to be treated 
with more care.

\section{Conclusions}

Canonical Hamiltonian field theory in curved spacetime 
has been  formulated in a manifestly covariant way, and 
quantization has been  achieved by a conventional correspondence 
principle. On a formal level, no problems related specifically 
 to the presence of gravity arise. In the case of the bosonic 
theory (which was assumed to be free of constraints), we obtained 
the expected result that the field momentum operator generates 
 spacetime translations on the field operators.  

On the other hand, in Dirac theory, we have to deal with second class 
constraints and the situation is less straightforward. 
We first performed a manifestly hermitian quantization, 
where it turned out that the field momentum does not correspond   
to the generator of spacetime translations, but rather to a modified 
translational operator, which has been identified, in our previous work, 
as a generalized, hermitian momentum operator. An alternative, not manifestly 
 hermitian quantization was achieved by modifying 
the  Lagrangian by a surface term. 
In this approach, the field momentum corresponds directly to the generator of 
translations, which is, however, given in terms of a non-hermitian Dirac 
operator. The change of the field momentum induced by the  surface term 
 was interpreted on a classical level 
as momentum transfer between the Dirac field and the gravitational background 
field. Both approaches should be physically equivalent. In order to show this 
on the quantum level, further investigations are necessary. It is expected 
that the addition of a surface term to the Lagrangian results in a 
change of the representation, such that both approaches can be related by 
a non-unitary transformation in operator space. Equivalence between 
both representations is then achieved by a simultaneous change of the 
scalar product in Hilbert space.

\section*{Acknowledgments}

This work has been supported by EPEAEK II in the framework of ``PYTHAGORAS 
II - SUPPORT OF RESEARCH GROUPS IN UNIVERSITIES'' (funding: 75\% ESF - 25\% 
National Funds).

\end{document}